\documentclass[12pt]{article}
\usepackage{a4wide,amssymb,cite}
\pdfoutput=1

\usepackage{a4wide,amssymb,graphicx}
\usepackage{epsfig}
\usepackage[usenames,dvipsnames]{color}
\usepackage{slashed}
\usepackage{amssymb,cite,graphicx}
\usepackage{slashed}
\usepackage{amsmath,bm,bbm}
\usepackage{amsfonts}
\usepackage[titletoc,title]{appendix}
\usepackage[small]{caption}
\usepackage[margin=1in]{geometry}
\usepackage[multiple]{footmisc}
\usepackage{mathtools}
\usepackage{slashed}
\usepackage{tabularx}
\usepackage[nottoc]{tocbibind}
\usepackage{xcolor}

\newcommand{\be}{\begin{equation}}
\newcommand{\ee}{\end{equation}}
\newcommand{\bea}{\begin{eqnarray}}
\newcommand{\eea}{\end{eqnarray}}

\begin{document}

\begin{titlepage}
%
%


%

\begin{centering}
\vspace{1cm}
{\Large {\bf Supersymmetry and LHC era}} \\

\vspace{1.5cm}

\begin{centering}
{\bf Hyun Min Lee$^{\dagger}$}
\end{centering}
\\
\vspace{.5cm}

{\it Department of Physics, Chung-Ang University, Seoul 06974, Korea.}

\vspace{.5cm}


\end{centering}
\vspace{1.5cm}

\begin{abstract}
\noindent
We review the basics of the supersymmetric extension of the Standard Model and discuss the implications of the constraints on the superpartner masses at the LHC for the Higgs mass, the $W$ boson mass, the muon $g-2$ and the proton lifetime.

\end{abstract}

\vspace{5cm}

{\it Invited review for Encyclopedia for Particle Physics, Elsevier}

\vspace{2cm}

\begin{flushleft} 
$^\dagger$Email: hminlee@cau.ac.kr 
\end{flushleft}

\end{titlepage}

\section{Introduction}\label{intro}

Supersymmetry(SUSY) for particle physics has been fostering a lot of interesting research avenues beyond the Standard Model (SM).
SUSY provides a superpartner for each particle in the SM, among the lightest of which can be a dark matter, solves the hierarchy problem in the SM, and helps the unification of gauge couplings at a high scale. We can test supersymmetric models by electroweak observables and  direct production of superpartners from the proton-proton collisions at the Large Hadron Collider (LHC). Since the first run at the LHC more than 15 years from now (as for May 2025) it is timely and worthwhile to summarize the status of the searches for superparticles at the LHC and ponder their implications for the recent anomalies such as the $W$ boson mass and the muon $g-2$, the unification of forces and the proton lifetime.

\section{Supersymmetric extension of the Standard Model}\label{sec1}

In the Minimal Supersymmetric Standard Model (MSSM), we introduce the chiral superfields for SM fermions and the Higgs doublet \footnote{We follow the notations and conventions from Wess and Bagger \cite{Wess:1992cp}.} as
\bea
{\hat Q}_i(y,\theta) &=& {\tilde q}_i(y) + \sqrt{2} \theta q_i (y)+ \theta^2 F_{Q_i}(y), \\
{\hat U}^c_i(y,\theta) &=& {\tilde u}^c_i (y)+ \sqrt{2} \theta u^c_i(y) + \theta^2 F_{U^c_i}(y), \\
{\hat D}^c_i(y,\theta) &=& {\tilde d}^c_i(y) + \sqrt{2} \theta d^c_i (y)+ \theta^2 F_{D^c_i}(y), \\
{\hat L}_i(y,\theta)&=& {\tilde l}_i(y) + \sqrt{2} \theta l_i(y) + \theta^2 F_{L_i}(y), \\
{\hat E}^c_i (y,\theta)&=&  {\tilde e}^c_i(y) + \sqrt{2} \theta e^c_i(y) + \theta^2 F_{E^c_i}(y), \\
{\hat H}_d(y,\theta) &=& H_d(y) +  \sqrt{2} \theta  {\tilde H}_d(y) + \theta^2 F_{H_d}(y)
\eea
where $\theta$ is the anti-commuting coordinate in components, $\theta^\alpha$ with $\alpha=1,2$, $\theta q_i\equiv \epsilon_{\alpha\beta}\theta^\alpha q^\beta_i$, etc, $\theta^2\equiv\epsilon_{\alpha\beta}\theta^\alpha\theta^\beta$ with $\epsilon_{\alpha\beta}$ being an antisymmetric $SL(2,C)$ invariant with $\epsilon_{12}=-1$, and $y^\mu=x^\mu+i\theta\sigma^\mu{\bar\theta}$.
Here, $u^c=(u^c)_L=(u_R)^c$, etc, and quark superpartners $ {\tilde q}_i ,  {\tilde u}^c_i,  {\tilde d}^c_i $ are squarks, lepton superpartners ${\tilde l}_i, {\tilde e}^c_i $  are sleptons, Higgs superpartner ${\tilde H}_d$ is Higgsino.  In order to cancel the $SU(2)_L\times U(1)_Y$ anomalies, we need to introduce one more Higgsino ${\tilde H}_u$, which makes an additional Higgs chiral multiplet,
\bea
{\hat H}_u(y,\theta) = H_u(y) +  \sqrt{2} \theta  {\tilde H}_u(y) + \theta^2 F_{H_u}(y). 
\eea
Therefore, there are two Higgs doublets in the MSSM.
All the chiral superfields satisfy the holomorphic constraints, ${\bar D}_{\dot\alpha}\Phi_I=0$, with $\Phi_I=\{{\hat Q}_i,{\hat U}^c_i, {\hat D}^c_i,{\hat L}_i,{\hat E}^c_i,{\hat H}_d,{\hat H}_u\}$, where  ${\bar D}_{\dot\alpha}=-\frac{\partial}{\partial {\bar\theta}^{\bar\alpha}}$ with ${\bar\theta}^{\dot\alpha}$ for ${\dot\alpha}=1,2$.
Then, the gauge-invariant superpotential is
\bea
W= y_{u,ij} {\hat Q}_i {\hat H}_u {\hat U}^c_j +  y_{d,ij} {\hat Q}_i {\hat H}_d {\hat D}^c_j +  y_{e,ij} {\hat L}_i {\hat H}_d {\hat E}^c_j  + \mu_H {\hat H}_u {\hat H}_d.
\eea

 There are spin-$\frac{1}{2}$ fermionic superpartners for the SM gauge bosons, called the gauginos, each of which is encoded in a vector superfield $V$.
 The vector superfield satisfies the reality condition $V=V^\dagger$, containing  the gauge field $v_\mu$, the gaugino $\lambda$ and the auxiliary field $D$, as follows,
\bea
V= -\theta \sigma^\mu {\bar\theta} v_\mu(x) + i\theta\theta {\bar\theta} {\bar\lambda}(x) -i {\bar\theta }{\bar\theta} \theta\lambda(x) +\frac{1}{2} \theta\theta {\bar\theta}{\bar\theta} \, D(x),
\eea
in Wess-Zumino gauge. Here, we also note that  ${\bar\theta}{\bar\lambda}\equiv\epsilon^{{\dot\alpha}{\dot\beta}} {\bar\theta}_{\dot\alpha}{\bar\lambda}_{\dot\beta}$ and ${\bar\theta}{\bar\theta}=\epsilon^{{\dot\alpha}{\dot\beta}}{\bar\theta}_{\dot\alpha}{\bar\theta}_{\dot\beta}$ with $\epsilon^{12}=1$. One can check that 
\bea
V^2= -\frac{1}{2}  \theta\theta {\bar\theta}{\bar\theta}\, v_\mu v^\mu,
 \eea
 and $V^3=0$.  
 Then, the superfield strength associated with the vector superfield is given by
 \bea
 W_\alpha &=&-i\lambda_\alpha(y) +\theta_\beta \Big[\delta^\beta_\alpha D(y)-\frac{i}{2} (\sigma^\nu {\bar\sigma}^\mu)_\alpha\,^\beta (\partial_\nu v_\mu(y)-\partial_\mu v_\nu(y)) \Big]  \nonumber \\
 &&+\theta\theta \sigma^\mu_{\alpha{\dot\alpha}} \partial_\mu {\bar\lambda}^{\dot\alpha}(y),
 \eea
In the MSSM, the gauginos include  ${\tilde g}^a$,  ${\tilde W}^i$ and ${\tilde B}$, called gluinos, Winos, and Bino, for the gauge bosons of $SU(3)_C\times SU(2)_L\times U(1)_Y$, respectively.

As a result, the gauge-invariant supersymmetric Lagrangian is given by
\bea
{\cal L}_{\rm MSSM} &=&\frac{1}{4} \Big[  {\rm Tr}(W^{\alpha} W_\alpha)|_{\theta\theta}+{\rm h.c.}\Big]+ \Phi^\dagger_I  e^{2 g V} \Phi_I|_{\theta\theta{\bar\theta}{\bar\theta}} + (W|_{\theta\theta} +{\rm h.c.}) \nonumber \\
&=&{\cal L}_{\rm cov-kin}+{\cal L}_{\rm Yukawa} +{\cal L}_{\rm gaugino} - V 
\eea
Here, ${\cal L}_{\rm cov-kin}$ stands for the covariant kinetic terms for the SM particles and their superpartners, the scalar potential $V$ is composed of $V=V_F+V_D$ where the the F-term potential is given by
\bea
V_F
&=& \Big|\frac{\partial W}{\partial Q_i} \Big|^2+ \Big|\frac{\partial W}{\partial U^c_j} \Big|^2+\Big|\frac{\partial W}{\partial D^c_j} \Big|^2 \nonumber \\
&&+ \Big|\frac{\partial W}{\partial L_i} \Big|^2+\Big|\frac{\partial W}{\partial E^c_j} \Big|^2+ \Big|\frac{\partial W}{\partial H_u} \Big|^2+\Big|\frac{\partial W}{\partial H_d} \Big|^2 \nonumber \\
&=&y^2_{u,ij} |H_u|^2(|{\tilde u}^c_j|^2 +|{\tilde q}_i|^2) +y^2_{d,ij} |H_d|^2(|{\tilde d}^c_j|^2 +|{\tilde q}_i|^2) + y^2_{e,ij} |H_d|^2(|{\tilde e}^c_j|^2 +|{\tilde l}_i|^2)\nonumber \\
&&+ |y_{u,ij}{\tilde q}_i {\tilde u}^c_j+\mu_H H_d|^2+   |y_{d,ij}{\tilde q}_i {\tilde d}^c_j+y_{e,ij}{\tilde l}_i {\tilde e}^c_j+\mu_H H_u|^2,
\eea
and the D-term potential is given by
\bea
V_D =\frac{1}{2} g^{\prime 2} \Big(\sum_I\phi^\dagger_I Y_I \phi_I\Big)^2 + \frac{1}{2} g^2 \bigg( \sum_I\phi^\dagger_I\frac{\tau^i}{2} \phi_I\bigg)^2 +\frac{1}{2}g^2_S \bigg( \sum_I\phi^\dagger_I\frac{\lambda^a}{2} \phi_I\bigg)^2,
\eea
with $Y_I, \tau^i, \lambda^a$ being hypercharge, Pauli and Gell-Mann matrices, respectively.
Moreover, the Yukawa couplings contain those in the SM as well as new interactions in the following,
\bea
-{\cal L}_{\rm Yukawa} &=& y_{u,ij} \Big(q_i H_u u^c_j + q_i {\tilde H}_u {\tilde u}^c_j+{\tilde q}_i {\tilde H}_u u^c_j \Big) \nonumber \\
&&+  y_{d,ij} \Big( q_i H_d d^c_j  +q_i {\tilde H}_d {\tilde d}^c_j+ {\tilde q}_i {\tilde H}_d d^c_j \Big) \nonumber \\
&&+  y_{e,ij} \Big(l_i H_d e^c_j +l_i {\tilde H}_d {\tilde e}^c_j+{\tilde l}_i {\tilde H}_d e^c_j \Big)  \nonumber \\
&&+ \mu_H {\tilde H}_u {\tilde H}_d+{\rm h.c.},
\eea
and the gaugino interactions are given by
\bea
{\cal L}_{\rm gaugino}&=&-\sqrt{2} g' \Big({\tilde l}^\dagger_{iL} Y_{l_L}  {\widetilde B} l_{iL} + {\tilde e}_{iR} Y_{e^c_R} {\widetilde B}  e^c_{iR}+{\rm h.c.} \Big)-\sqrt{2}g \bigg({\tilde l}^\dagger_{iL}  \frac{\tau^3}{2} {\widetilde W}^3 l_{iL} +{\rm h.c.}\bigg) \nonumber \\
&&-g \Big({\tilde e}^*_{iL} {\widetilde W}^- \nu_{iL} +{\tilde\nu}^*_{iL} {\widetilde W}^+ e_{iL} +{\rm h.c.} \Big)\nonumber \\
&&+{\rm squark\,\,  interactions}.
\eea
Here, we wrote the leptons and sleptons as $l_{iL}=(\nu_{iL},e_{iL})^T$, ${\tilde l}_{iL}=({\tilde\nu}_{iL},{\tilde e}_{iL})^T$ and $e^c_{iR}$ and ${\tilde e}^*_{iR}$.
We note that the Higgsinos have a Dirac mass $\mu_H$ and the Higgs doublets have the same masses as the Higgsino mass by SUSY.
If there is a chiral symmetry under which Higgsinos are charged, the Higgsino mass is naturally small by chiral symmetry, so the small masses of Higgs doublets are ensured by SUSY.

The SUSY algebra is extended by the R-symmetry generator, which does not commute with the SUSY operators,
\bea
[Q_\alpha,R]=-Q_\alpha, \quad\quad [{\bar Q}_{\dot\alpha},R]= + {\bar Q}_{\dot\alpha}.
\eea
The superfield $\Phi=\phi+\sqrt{2} \theta \psi +\theta^2 F_\phi$ has the same $R$-charge as the boson, so the anti-commuting coordinate $\theta$ has $R$-charge $+1$, so the auxiliary field $F_\phi$ has $R$-charge $r_F=r_B-2$. The $R$-symmetry transformation for a chiral superfield $\Phi$ with $R$-charge $r$ is
\bea
R \,\Phi(\theta,x)&=& e^{ir\alpha} \Phi(e^{-i\alpha}\theta,x), \\
R\, \Phi^\dagger({\bar\theta},x)&=& e^{ir\alpha} \Phi^\dagger(e^{i\alpha}{\bar\theta},x),
\eea
or
\bea
\phi&\longrightarrow& e^{ir\alpha} \phi, \\
\psi & \longrightarrow & e^{i(r-1)\alpha} \psi, \\
F_\phi &\longrightarrow & e^{i(r-2)\alpha} F_\phi.
\eea
On the other hand, vector superfields have zero $R$-charge, because they are real, so $R\, V(\theta, {\bar\theta}, x)=V(e^{-i\alpha}\theta, e^{i\alpha} {\bar\theta},x)$. 
So, in Wess-Zumino gauge, the $R$-transformations for components are
\bea
v_\mu\rightarrow v_\mu,\quad \lambda\rightarrow e^{i\alpha} \lambda, \quad D\rightarrow D.
\eea
The $R$-parity is a $Z_{2R}$ discrete symmetry with $\alpha=\pi$.
In MSSM, we take $r=1$  for quark and lepton chiral multiplets and $r=0$ for Higgs chiral multiplets. Then, the $R$-parities for chiral superfields in MSSM are assigned as
\bea
{\hat Q}_i, \,\,{\hat U}^c_i, \,\,{\hat D}^c_i,\,\, {\hat L}_i,\,\, {\hat E}^c_i &:&  \quad Z_{2R}=-1, \\
{\hat H}_{u,d} &:& \quad Z_{2R}=+1.
\eea
so the $R$-parities for component fields are
\bea
q_i, \,\, u^c_i,  \,\, d^c_i, \,\, l_i, \,\, e^c_i, \,\, H_{u,d}, \,\, v_{m,i} &:& \quad Z_{2R}=+1, \\
{\tilde q}_i, \,\,  {\tilde u}^c_i, \,\,  {\tilde d}^c_i, \,\, {\tilde l}_i, \,\, {\tilde e}^c_i, \,\, {\tilde H}_{u,d}, \,\, \lambda_i &:&  \quad Z_{2R}=-1.
\eea
The $R$-parity is related to the matter parity $P_M=(-1)^{3(B-L)}$ by $Z_{2R}=(-1)^{2S}P_M$  with $S$ being the spin.
Therefore, the Lightest Supersymmetric Particle (LSP) among the neutral components of ${\tilde W}^i$, ${\tilde H}_u$, ${\tilde H}_d$, and ${\tilde B}$ (neutralinos), is a good candidate for WIMP dark matter.  The stability of LSP is ensured by $R$-parity, which is a global symmetry in MSSM as discussed above. The $R$-parity can be generalized to discrete $R$ symmetries such as $Z_{4R}$, $Z_{6R}$, etc. In this case, the $\mu$-problem and the proton decay problem in the minimal $SU(5)$ GUTs can be solved automatically \cite{Lee:2010gv,Lee:2011dya}.

If SUSY is exact, each superpartner would have the same mass as its partner particle in the SM. 
Thus, SUSY is broken in nature,  so we need to make superparticles heavier than the SM counterparts while maintaining the cancellation of quadratic divergences for the scalar fields in the MSSM.
To this, we introduce soft SUSY breaking terms,
\bea
{\cal L}_{\rm soft}&=& -\Big(\frac{1}{2} \sum^3_{i=1} M_i \lambda_i \lambda_i +{\rm h.c.}\Big) -m^2_{H_d}|H_d|^2- m^2_{H_u}|H_u|^2 \nonumber \\
&&- m^2_{{\tilde q}, ij} |{\tilde q}_i|^2- m^2_{{\tilde u}^c, ij} |{\tilde u}^c_i|^2- m^2_{{\tilde d}^c, ij} |{\tilde u}^c_i|^2- m^2_{{\tilde l}, ij} |{\tilde l}_i|^2- m^2_{{\tilde e}^c, ij} |{\tilde e}^c_i|^2 \nonumber \\
&&-  T_{u,ij} {\tilde q}_i H_u {\tilde u}^c_j - T_{d,ij} {\tilde q}_i H_d {\tilde d}^c_j -T_{e,ij} {\tilde l}_i H_d {\tilde e}^c_j  +{\rm h.c.}.
\eea
For no FCNC, we usually choose soft masses to be flavor diagonal  by $m^2_{{\tilde q},ij}=m^2_{\tilde q} \, \delta_{ij}$, etc, and aligned by $T_{u,ij}=y_{u,ij} A_t$, etc. 

\begin{figure}[t]
	\centering
	\includegraphics[width=8cm,height=5cm]{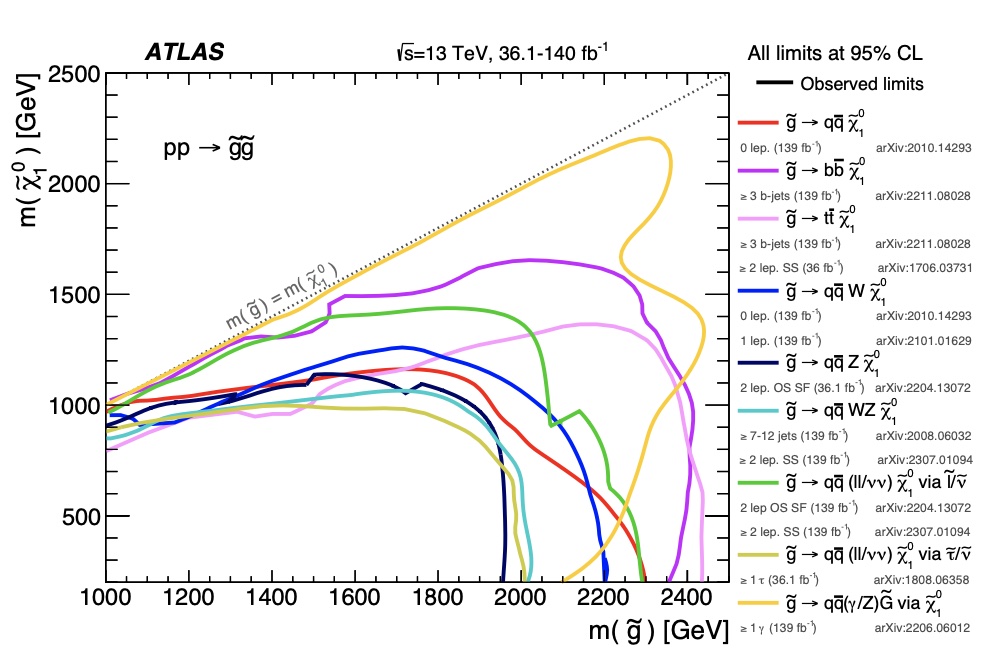} \,\,\,\,
	\includegraphics[width=7cm,height=5cm]{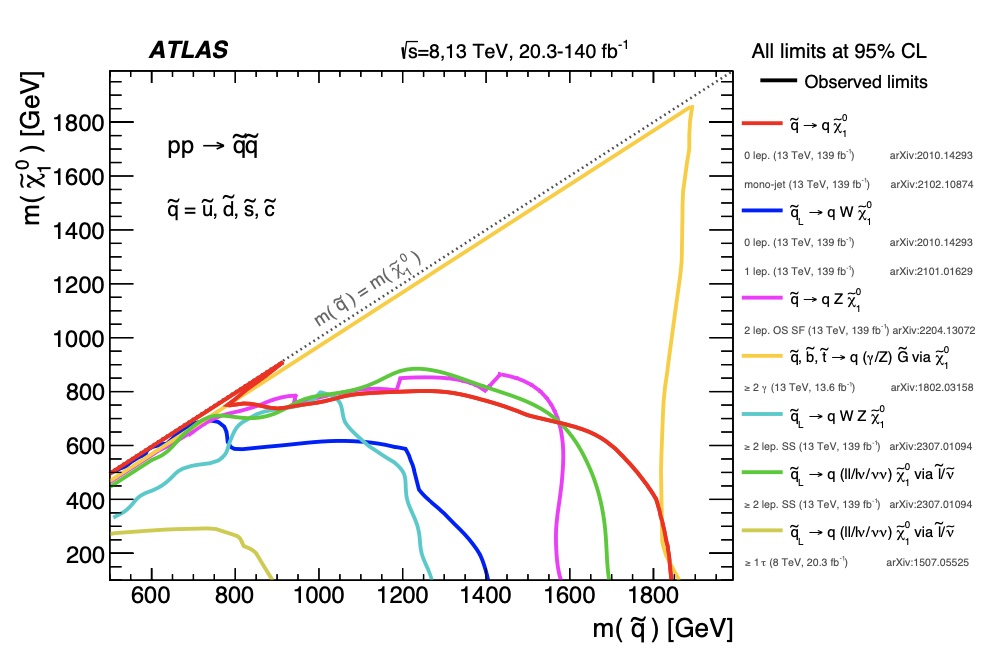}
	\caption{Limits on gluino and squark masses from ATLAS in the left and right plots, respectively, extracted from Ref.~\cite{ATLAS:2024lda}.}
	\label{fig:gluino}
\end{figure}

In Fig.~\ref{fig:gluino}, we show that various searches at the LHC have set the bounds on the gluino mass in the left plot and the squark mass in the right plot, respectively, relative to the lightest neutralino mass. Thus, the gluino masses up to about $2.4\,{\rm TeV}$ and the squark masses up to about $1.8\,{\rm TeV}$ have been ruled out by ATLAS \cite{ATLAS:2024lda}. Similar limits from CMS \cite{CMS:2019zmd,CMS:2020bfa} can be also found.

\section{Supersymmetry and Higgs potential}

Weak-scale SUSY has been motivated to explain the hierarchy problem in the SM, because the superpartners with weak-scale masses cancel the huge loop contributions to the Higgs mass parameter so they maintain the electroweak scale as compared to the ultraviolet scale such as the Planck scale as observed.

First, we recall the SM Higgs potential, which takes a simple form as
\bea
V=m^2_H |H|^2 +\lambda_H |H|^4.
\eea
Then, after electroweak symmetry breaking with the VEV of the Higgs field,
\bea
v=\sqrt{2}\langle |H| \rangle = \sqrt{-\frac{m^2_H}{\lambda_H}}.
\eea
$W$ and $Z$ bosons receive masses,
\bea
m^2_W = \frac{1}{4} g^2 v^2,\quad m^2_Z = \frac{1}{4} (g^2+g^{\prime 2}) v^2.
\eea
Here, the Higgs VEV is determined to be $v=246\,{\rm GeV}$ by the measurement of the Fermi constant with the following relation,
\bea
\frac{G_F}{\sqrt{2}} = \frac{g^2}{8m^2_W} =\frac{1}{2v^2}
\eea
where $G_F=1.16639\times 10^{-5}\,{\rm GeV}^{-2}$ is  given by the muon decay, $\mu\rightarrow e {\bar\nu}_e \nu_\mu$. 
On the other hand, from the expansion around the Higgs vacuum, $H=\frac{1}{\sqrt{2}}(0,v+h)^T$, the Higgs boson mass determines the Higgs mass parameter by
\bea
m_h=\sqrt{2\lambda_H} \, v = \sqrt{2}|m_H|=125\,{\rm GeV}. \label{weakscale}
\eea

The top Yukawa coupling is the strongest in the SM, contributing most to the Higgs mass parameter at loop level.  In the decoupling limit of the extra Higgs fields for $H^0_u=\frac{1}{\sqrt{2}} h \sin\beta$ and $H^0_d=\frac{1}{\sqrt{2}} h \cos\beta$ where $\tan\beta=\frac{\langle H^0_u\rangle}{\langle H^0_d\rangle}$, the relevant couplings for the Higgs mass corrections are
\bea
{\cal L}_{\rm MSSM}\supset-\Big( \frac{1}{\sqrt{2}}\ y_t \, h {\bar t}_L  t_R+{\rm h.c.}\Big) - \frac{1}{2} y^2_t h^2(|{\tilde t}_R|^2 +|{\tilde t}_L|^2)
\eea
where ${\tilde t}_R= ({\tilde t}^c)^*$ and $y_t=y_{u,33}\sin\beta$. 
Then, the top loop corrections to the Higgs mass parameter are
\bea
(\Delta m^2_{H})_t=-\frac{N_c y^2_t}{8\pi^2}\, \Lambda^2+ \frac{3N_c y^2_t}{8\pi^2}\, m^2_t \ln \Big(\frac{\Lambda}{m_t} \Big).
\eea
whereas the stop loop contributions to the Higgs mass parameter are
\bea
(\Delta m^2_{H})_{\tilde t}=\frac{N_c y^2_t}{8\pi^2}\, \Lambda^2- \frac{N_c y^2_t}{8\pi^2}\, m^2_{\tilde t} \ln \Big(\frac{\Lambda}{m_{\tilde t}} \Big).
\eea
Here, $\Lambda$ is the physical cutoff of the theory such as the Planck scale in gravity mediation and the GUT scale in gauge mediation.
As a result, adding both top and stop contributions, the quadratic divergences are cancelled out, but the modified Higgs mass parameter become
\bea
\Delta m^2_{H}=- \frac{N_c y^2_t}{8\pi^2}\, m^2_{\tilde t} \ln \Big(\frac{\Lambda}{m_{\tilde t}} \Big), \quad m_{\tilde t}\gg m_t,
\eea
due to the non-discovery of the stops at the LHC, raising the problem of a fine-tuning for the Higgs mass parameter to the weak scale as required by eq.~(\ref{weakscale}).
However, one could resort the solution for the little hierarchy problem in the MSSM to various proposals such as composite Higgs models, clockwork models, twin Higgs models, relaxion models, four-form flux models, etc \cite{Lee:2019zbu}.

On the other hand, the quartic terms for neutral Higgs fields come from the D-term potential as follows,
\bea
V= \frac{1}{8} (g^2+ g^{\prime 2}) (|H^0_u|^2- |H^0_d|^2)^2  
\eea 
Then, in the decoupling limit of the extra Higgs fields, we obtain the quartic terms as
\bea
V=\frac{1}{32} (g^2+ g^{\prime 2}) \cos^2(2\beta) \, h^4=\frac{1}{4} \lambda_H h^4. 
\eea
Thus, the Higgs quartic coupling is given by the electroweak gauge couplings as
\bea
\lambda_H = \frac{1}{8} (g^2+ g^{\prime 2}) \cos^2(2\beta).
\eea
Therefore, as far as superparticle masses are below the vacuum instability scale, the Higgs quartic coupling is maintained to be positive all the way to the unification scale. From the quartic coupling at tree level, the Higgs boson mass  is given by
\bea
m_h=\sqrt{2\lambda_H} \, v = \frac{1}{2} \sqrt{g^2+ g^{\prime 2}} \cos(2\beta) \, v \leq m_Z,
\eea
which is too small to explain the Higgs mass.
Thus, after including the top loop corrections to the Higgs quartic coupling, we can accommodate the correct Higgs boson mass with a shift in the quartic coupling,
\bea
\Delta\lambda_H = \frac{3m^4_t}{4\pi^2 v^4}\, \bigg[\ln \Big(\frac{m^2_{\tilde t}}{m^2_t}\Big) + \frac{X^2_t}{m^2_{\tilde t}}\Big( 1-\frac{1}{12}  \frac{X^2_t}{m^2_{\tilde t}}\Big) \bigg]
\eea
with $X_t=A_t -\mu_H \cot\beta$. In this case, the required stop masses are required to be at least multi-TeV scales.

\begin{figure}[t]
	\centering
	\includegraphics[width=8cm,height=5cm]{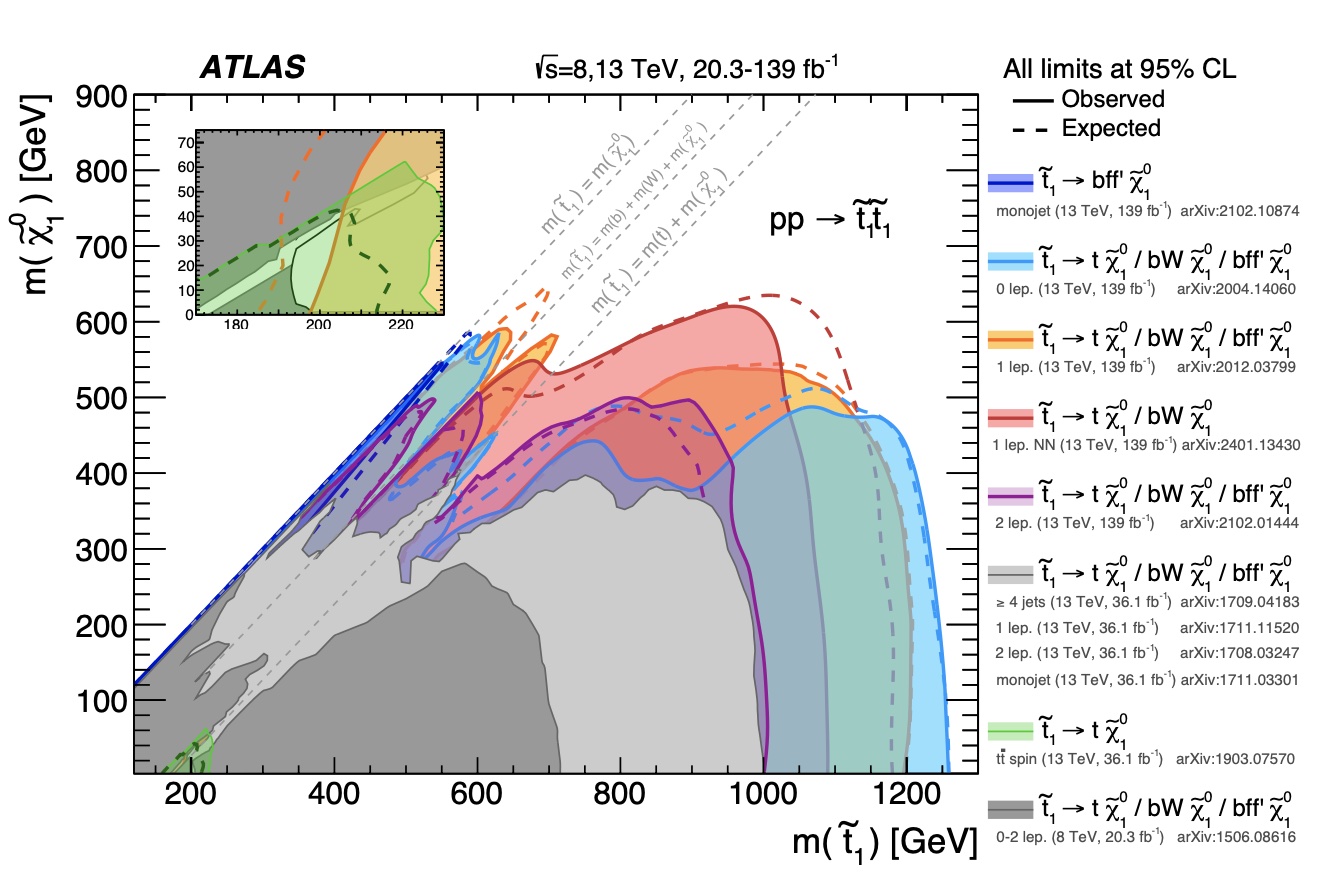} \,\,\,\,
	\includegraphics[width=7cm,height=5cm]{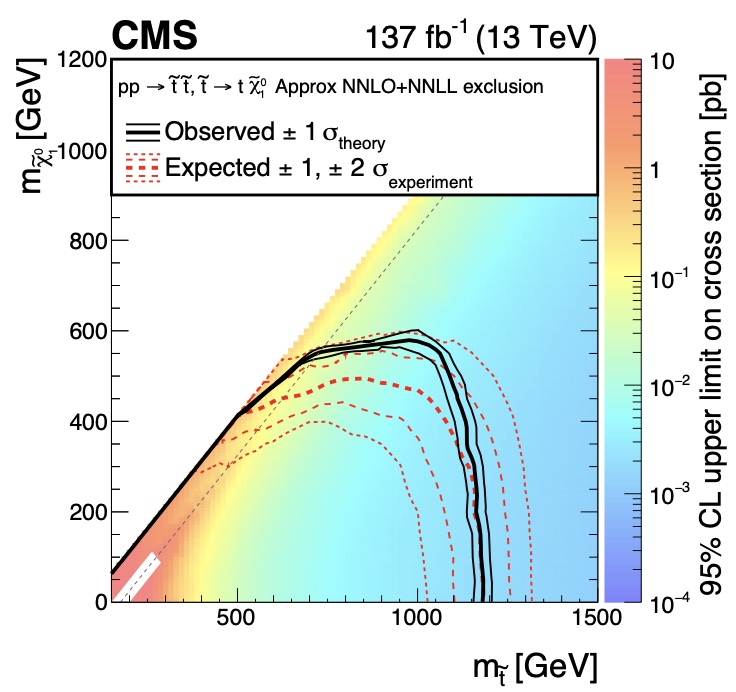} 
	\caption{Limits on the stop mass vs the neutralino mass at LHC, extracted from Refs.~\cite{ATLAS:2024lda} and \cite{CMS:2019zmd}.}
	\label{fig:stop}
\end{figure}

In Fig.~\ref{fig:stop}, the current limits on the stop mass vs the neutralino mass at the LHC, depending on the decay modes of the stop, showing that the stop masses up to about $1250\,{\rm GeV}$ have been ruled out from ATLAS data \cite{ATLAS:2024lda} and there are similar bounds on the stop mass from CMS \cite{CMS:2019zmd}.

\section{Supersymmetry, $W$ boson mass and muon $g-2$}

Superpartners of left and right-handed fermions have mixing mass terms, and electroweak gauginos and Higgsinos also have mixing mass terms after electroweak symmetry breaking. We focus on the slepton contributions to the  $W$ boson mass and the muon $g-2$.

\subsection{Sleptons and $W$ boson mass}

First, we can derive the theoretical value of the $W$ boson mass from the muon decay amplitude, which relates $M_W$ to the Fermi constant $G_\mu$, the fine structure constant $\alpha$, and the $Z$ boson mass $M_Z$, as follows,
\bea
M^2_W \bigg( 1-\frac{M^2_W}{M^2_Z}\bigg) =\frac{\pi \alpha}{\sqrt{2} G_\mu}  ( 1+ \Delta r ) 
\eea
where $\Delta r$ encodes the loop corrections in the SM and the contributions from new physics.
We note that $\Delta r=0.0381$ in the SM, which leads to the SM prediction for the $W$ boson mass, as follows \cite{Haller:2018nnx,ParticleDataGroup:2024cfk}, 
\bea
M^{\rm SM}_W=80.353\,{\rm GeV}\pm 6\,{\rm MeV}. 
\eea
In comparison, the world average for the measured $W$ boson mass  in PDG, including ATLAS but without CDFII  \cite{ParticleDataGroup:2024cfk}, is given by 
\bea
M^{\rm PDG}_W= 80.366\,{\rm GeV}\pm 12\,{\rm MeV}.
\eea
Thus, the SM prediction for the $W$ boson mass is consistent with the PDG value within $2\sigma$.
On the other hand, the Fermilab CDFII experiment  \cite{CDF:2022hxs} has recently measured the $W$ boson mass as
\bea
M^{\rm CDFII}_W= 80.4335\,{\rm GeV}\pm 9.4\,{\rm MeV}. \label{dw}
\eea
So, if confirmed,  the result could show a considerable deviation from the SM prediction at the level of $7.0\sigma$, calling for a new physics explanation if confirmed. However, both ATLAS \cite{ATLAS:2024erm} and CMS \cite{CMS:2024lrd} at LHC have recently announced the measured values of $M_W$, 
\bea
M^{\rm ATLAS}_W&=& 80.3665\,{\rm GeV}\pm 15.9\,{\rm MeV}, \\
M^{\rm CMS}_W&=& 80.3602\,{\rm GeV}\pm 9.9\,{\rm MeV},
\eea
which are consistent with the SM predictions (with similar uncertainty in CMS as in CDFII). 

In the MSSM, the soft SUSY breaking masses are $SU(2)_L$ invariant, but the  $SU(2)_L$ breaking mass terms split between the masses of the scalar superpartners within the same $SU(2)_L$ doublet. Focusing on the sleptons \footnote{See Ref.~\cite{Lee:2012sy} for the contributions of stop and sbottom superpartners to the $\rho$ parameter in relation to the Higgs mass measurement.},  we obtain a sizable contribution from slepton loops to the $\rho$ parameter \cite{Heinemeyer:2004gx}, as follows,
\bea
\Delta \rho=\frac{3G_\mu}{8\sqrt{2}\pi^2} \bigg[-\sin^2\theta_{\tilde\mu}\cos^2\theta_{\tilde\mu} F_0(m^2_{{\tilde\mu}_1},m^2_{{\tilde\mu}_2})
+\cos^2\theta_{\tilde\mu} F_0(m^2_{\tilde\nu},m^2_{{\tilde\mu}_2})+\sin^2\theta_{\tilde\mu} F_0(m^2_{\tilde\nu},m^2_{{\tilde\mu}_1})\bigg]
\eea
with
\bea
F_0(x,y)=x+y -\frac{2xy}{x-y}\ln \frac{x}{y}.
\eea
As the new physics contribution is related to the correction to the $\rho$ parameter by 
\bea
(\Delta r)_{\rm new} = -\frac{c^2_W}{s^2_W}\,\Delta\rho, \label{deltar}
\eea
we can turn the correction to the $\rho$ parameter into the correction to the W boson mass by
\bea
\Delta M_W \simeq \frac{1}{2} M_W\,\frac{c^2_W}{c^2_W-s^2_W}\, \Delta \rho. \label{Wmass}
\eea
We note that the global fit in PDG without CDFII is given by $\Delta\rho=(0.0\pm 5.0)\times 10^{-4}$ for $U=0$ \cite{ParticleDataGroup:2024cfk}, which can constrain the slepton masses and mixings.

\subsection{Sleptons and muon $g-2$}

The value of the muon $g-2$ measured by Brookhaven experiment \cite{Muong-2:2006rrc}  has been recently confirmed by Fermilab experiment \cite{Muong-2:2021ojo,Muong-2:2023cdq}. Thus, the values of the muon $g-2$ from the experiment and the SM prediction are given by
\bea
a^{\rm exp}_\mu &=&116592059(22)\times 10^{-11}, \\
a^{\rm SM}_\mu &=& 116591810(1)(40)(18)\times 10^{-11},
\eea
respectively,
exhibiting the deviation from the SM predictions \cite{Aoyama:2020ynm} at $5.2\sigma$ from $\Delta a_\mu\equiv a^{\rm exp}_\mu-a^{\rm SM}_\mu=249(22)(43)\times 10^{-11}$ \cite{ParticleDataGroup:2024cfk}.

However, the SM predictions for the muon $g-2$ based on the lattice results \cite{Borsanyi:2020mff,Boccaletti:2024guq} are consistent with its experimental value, and the recent CMD-3 data \cite{CMD-3:2023rfe} shows a sizable deviation from the other $e^+e^-$ data, which were used to derive the SM prediction for the muon $g-2$ in the dispersive approach \cite{Aoyama:2020ynm}. Thus, it is crucial to understand the hadronic contributions to the muon $g-2$ well within the SM from the lattice and experimental data. Nonetheless, it is still illuminating to investigate what the impact of supersymmetric particles is on the muon $g-2$ \cite{Endo:2021zal,Kim:2024bub}.

The mass terms for neutralinos, $\{{\widetilde B}, {\widetilde W}^3,{\widetilde H}^0_{d}, {\widetilde H}^0_{u}\}$, charginos, $\{{\widetilde W}^-, {\widetilde H}^-_d\}$,  and their complex conjugates, are given, respectively, by
\bea
{\cal L}_N=-\frac{1}{2} ({\widetilde B}, {\widetilde W}^3,{\widetilde H}^0_{d}, {\widetilde H}^0_{u}){\cal M}_N \left(\begin{array}{c} {\widetilde B} \\{\widetilde W}^3 \\  {\widetilde H}^0_d \\ {\widetilde H}^0_u \end{array} \right)+{\rm h.c.}
\eea
with
\bea
{\cal M}_N =\left(\begin{array}{cccc} M_1 &  0 & -m_Z\sin\theta_W \cos\beta & m_Z\sin\theta_W\sin\beta \\ 0  & M_2  & m_Z\cos\theta_W\cos\beta  & -m_Z\cos\theta_W\sin\beta \\  -m_Z\sin\theta_W \cos\beta & m_Z\cos\theta_W\cos\beta & 0  & -\mu_H  \\ m_Z\sin\theta_W \sin\beta & -m_Z\cos\theta_W\sin\beta  & -\mu_H & 0 \end{array} \right),
\eea
and
\bea
{\cal L}_C=-(\overline{{\widetilde W}^-_R}, \overline{{\widetilde H}^-_{uR}} ) {\cal M}_C \left(\begin{array}{c}{\widetilde W}^-_L  \\ {\widetilde H}^-_{dL} \end{array}\right)
\eea
with
\bea
{\cal M}_C=\left(\begin{array}{cc}  M_2 & \sqrt{2} m_W \cos\beta \\ \sqrt{2}m_W \sin\beta  & \mu_H  \end{array}\right).
\eea
Here, $M_{1,2}$ are the soft SUSY breaking masses for Bino and Wino gauginos, and $\mu_H$ is the supersymmetric mass parameter for the Higgsinos. We note that $\sin\beta=v_u/v$ and $\cos\beta=v_d/v$ with $v=\sqrt{v^2_u+v^2_d}$,  for $\langle H^0_u\rangle=\frac{1}{\sqrt{2}}v_u$ and  $\langle H^0_d\rangle=\frac{1}{\sqrt{2}}v_d$.

The mass matrix for neutralinos can be diagonalized by
\bea
N^*{\cal M}_N N^\dagger = {\cal M}^{\rm diag}_N= {\rm diag}(m_{{\widetilde \chi}^0_1},m_{{\widetilde \chi}^0_2},m_{{\widetilde \chi}^0_3},m_{{\widetilde \chi}^0_4})
\eea
where the rotation matrix $N$ defines the mass eigenstates for neutralinos in four-component spinor notations as
\bea
{\widetilde \chi}^0_{iL} &=& N_{i1} {\widetilde B}_L+N_{i2} {\widetilde W}^3_L+ N_{i3} {\widetilde H}^0_{d,L}+N_{i4} {\widetilde H}^0_{u,L}, \\
{\widetilde \chi}^0_{iR} &=& N^*_{i1} {\widetilde B}_R+N^*_{i2} {\widetilde W}^3_R+ N^*_{i3} {\widetilde H}^0_{d,R}+N^*_{i4} {\widetilde H}^0_{u,R}.\label{neutralinomix}
\eea
In the limit of small electroweak symmetry breaking effects, namely, $m_Z\ll |\mu_H\pm M_1|, |\mu\pm M_2|$, we can approximate the neutralino masses as
\bea
m_{{\widetilde \chi}^0_1} &\simeq & M_1 -\frac{m^2_Z s^2_W (M_1+\mu_H\sin 2\beta)}{\mu^2_H-M^2_1}, \\
m_{{\widetilde \chi}^0_2} &\simeq & M_2-\frac{m^2_W (M_2+\mu_H\sin 2\beta)}{\mu^2_H-M^2_2}, \\
m_{{\widetilde \chi}^0_3} &\simeq & \mu_H+ \frac{m^2_Z(1+\sin2\beta)(\mu_H-M_1 c^2_W-M_2s^2_W)}{2(\mu_H-M_1)(\mu_H-M_2)}, \\
m_{{\widetilde \chi}^0_4} &\simeq & -\mu_H  + \frac{m^2_Z(1-\sin2\beta)(\mu_H+M_1 c^2_W+M_2s^2_W)}{2(\mu_H+M_1)(\mu_H+M_2)}.
\eea
Here, for $\mu_H>0$, the leading mass for ${\widetilde \chi}^0_4$ is negative, so we need to rescale ${\widetilde \chi}^0_4$ to $i{\widetilde \chi}^0_4$ to get a positive mass for ${\widetilde \chi}^0_4$. On the other hand, for $\mu_H<0$, instead we need to rescale ${\widetilde \chi}^0_3$ to $i{\widetilde \chi}^0_3$ to get a positive mass for ${\widetilde \chi}^0_3$.
Moreover, the components of the neutralino mixing matrix in eq.~(\ref{neutralinomix}) are approximated to
{\small
\bea
N_{11}=1, \,\, N_{12}=0,  \,\,
N_{13}= \frac{(M_1\cos\beta+\mu_H \sin\beta)m_Z s_W}{\mu^2_H-M^2_1},  \,\,
N_{14}= - \frac{(M_1\sin\beta+\mu_H \cos\beta)m_Z s_W}{\mu^2_H-M^2_1},
\eea
}
{\small
\bea
N_{21}=0, \,\, N_{22}=1, \,\,
N_{23}=- \frac{(M_2\cos\beta+\mu_H \sin\beta)m_Z c_W}{\mu^2_H-M^2_2}, \,\,
 N_{24}= \frac{(M_2\sin\beta+\mu_H \cos\beta)m_Z c_W}{\mu^2_H-M^2_2},
\eea
}
{\small
\bea
N_{31} = -\frac{1}{\sqrt{2}}\,\frac{m_Zs_W}{\mu_H-M_1}\,(\cos\beta+\sin\beta), \,\,
N_{32} = \frac{1}{\sqrt{2}}\,\frac{m_Zc_W}{\mu_H-M_2}\,(\cos\beta+\sin\beta), \,\,
N_{33}=-N_{34}=\frac{1}{\sqrt{2}},
\eea
}
{\small
\bea
N_{41} = \frac{1}{\sqrt{2}}\,\frac{m_Z s_W}{\mu_H+M_1}\,(\cos\beta-\sin\beta), \,\,
N_{42} = -\frac{1}{\sqrt{2}}\,\frac{m_Z c_W}{\mu_H+M_2}\,(\cos\beta-\sin\beta), \,\,
N_{43}=N_{44}=\frac{1}{\sqrt{2}}.
\eea
}
Thus, in the leading order approximation with a heavy Higgsino, the neutralino mass eigenstates become ${\widetilde\chi}^0_1\simeq {\widetilde B}$ (bino-like),  ${\widetilde\chi}^0_2\simeq {\widetilde W}^3$ (wino-like), ${\widetilde\chi}^0_3, {\widetilde\chi}^0_4 \simeq ({\widetilde H}^0_u\mp  {\widetilde H}^0_d)/\sqrt{2}$ (Higgsino-like). 

Similarly, the mass matrix for charginos can be also diagonalized by
\bea
U_R {\cal M}_C U^\dagger_L ={\cal M}^{\rm diag}_C= {\rm diag}(m_{{\widetilde \chi}^-_1},m_{{\widetilde \chi}^-_2}), 
\eea
with
{\small
\bea
m^2_{{\widetilde \chi}^-_1,{\widetilde \chi}^-_2} =\frac{1}{2}\bigg[|M_2|^2+|\mu_H|^2+2m^2_W \mp \sqrt{(|M_2|^2+|\mu_H|^2+2m^2_W)^2-4|\mu_H M_2-m^2_W\sin 2\beta|^2} \bigg],
\eea
}
and the mass eigenstates for charginos are
\bea
 \left(\begin{array}{c}{\widetilde \chi}^-_1  \\ {\widetilde \chi}^-_2 \end{array}\right)_{L,R} =U_{L,R}  \left(\begin{array}{c}{\widetilde W}^-_L  \\ {\widetilde H}^-_{dL} \end{array}\right)_{L,R}. \label{charginomix}
\eea
In the limit of $m_W\ll |\mu_H\pm M_2|$, the chargino masses are approximated to
\bea
m_{{\widetilde \chi}^-_1} &\simeq&M_2 -\frac{m^2_W (M_2+\mu_H\sin 2\beta)}{\mu^2_H-M^2_2},  \\
m_{{\widetilde \chi}^-_2} &\simeq&\mu_H+ \frac{m^2_W (\mu_H+M_2\sin 2\beta)}{\mu^2_H-M^2_2},
\eea
and chargino mass eigenstates become ${\widetilde \chi}^-_1 \simeq {\widetilde W}^-$ (wino-like), ${\widetilde \chi}^-_2 \simeq {\widetilde H}^-_d$ (higgsino-like).  For $\mu_H<0$, we need to rescale the Higgsino-like chargino by ${\widetilde \chi}^-_2\to i{\widetilde \chi}^-_2$ to get a positive mass for ${\widetilde \chi}^-_2$. We refer to Appendix A for the neutralino and chargino mixing matrices with corrections coming from the electroweak symmetry breaking. 
Moreover, the components of the chargino mixing matrices in eq.~(\ref{charginomix}) are approximated to
\bea
U_{L,11}=U_{L,22}=1, \quad
U_{L,12}=-U_{L,21}=-\frac{\sqrt{2}m_W(M_2\cos\beta+\mu_H \sin\beta)}{\mu^2_H-M^2_2},
\eea
\bea
U_{R,11}=U_{R,22}=1, \quad
U_{R,12}=-U_{R,21}=-\frac{\sqrt{2}m_W(M_2\sin\beta+\mu_H \cos\beta)}{\mu^2_H-M^2_2}.
\eea

The mass terms for the charged sleptons are given by
\bea
{\cal L}_{\tilde e} = -({\tilde e}^*_L, {\tilde e}^*_R) {\cal M}^2_{\tilde e} \left(\begin{array}{c} {\tilde e}_L \\ {\tilde e}_R \end{array}\right), \quad {\cal M}^2_{\tilde e}=\left(\begin{array}{cc} m^2_{LL} & m^2_{LR} \\  (m^2_{LR})^* & m^2_{RR} \end{array}\right) \label{sleptonmass}
\eea
where 
\bea
m^2_{LL}&=&m^2_{{\tilde e}_L}+m^2_Z\cos2\beta \Big(s^2_W-\frac{1}{2}\Big)+m^2_e,  \\
m^2_{RR}&=&m^2_{{\tilde e}_R}-m^2_Z\cos2\beta \,s^2_W+m^2_e, \\
m^2_{LR}&=& m_l (A^*_l -\mu_H \tan\beta). 
\eea
After diagonalizing the slepton mass matrix in eq.~(\ref{sleptonmass}) with
\bea
 \left(\begin{array}{c} {\tilde e}_2\\ {\tilde e}_1 \end{array}\right) =\left(\begin{array}{cc} \cos\theta_{\tilde e} & \sin\theta_{\tilde e} \\ -\sin\theta_{\tilde e} & \cos\theta_{\tilde e} \end{array}\right) \left(\begin{array}{c} {\tilde e}_L \\ {\tilde e}_R \end{array}\right),  \label{sleptonmix}
\eea
we obtain the slepton mass eigenvalues as
\bea
m^2_{{\tilde e}_2, {\tilde e}_1} = \frac{1}{2} \bigg[m^2_{LL}+m^2_{RR} \pm (m^2_{LL}-m^2_{RR})\sqrt{1+\frac{4|m^2_{LR}|^2}{(m^2_{LL}-m^2_{RR})^2}} \bigg],
\eea
and the slepton mixing angle as
\bea
\tan2\theta_{\tilde e} =\frac{2m^2_{LR}}{m^2_{LL}-m^2_{RR}},
\eea
or
\bea
\sin2\theta_{\tilde e} = \frac{2m^2_{LR}}{m^2_{{\tilde e}_2}-m^2_{{\tilde e}_1}}, \quad -\frac{\pi}{2}<\theta_{\tilde e}<\frac{\pi}{2}. \label{sleptonmix}
\eea

There are two typos of supersymmetric interactions relevant for the muon $g-2$.
One Yukawa type interaction of the muon is to a neutral scalar $\phi$ with mass $m_\phi$ and a charged fermion $F$ with charge $-1$ and mass $m_F$, given by
\bea
{\cal L}_1 = - {\bar \mu} \Big(A_S + A_P \gamma^5 \Big)F\phi +{\rm h.c.}. \label{gen1}
\eea
In this case,  the one-loop contribution to the muon $g-2$ is given \cite{Kim:2024bub} by
\bea
a^{(1)}_\mu &=&- \frac{m^2_\mu}{8\pi^2}\int^1_0 dx\, \frac{|A_S|^2 \big(x^3-x^2-x^2\,\frac{m_F}{m_\mu}\big)+|A_P|^2 (m_F\rightarrow -m_F)}{(1-x)m^2_\phi+(m^2_F-m^2_\mu)x+x^2 m^2_\mu} \nonumber \\
&=& \frac{m_\mu}{8\pi^2} \bigg(\frac{m_\mu}{12m^2_\phi} (|A_S|^2+|A_P|^2) F^C_1(x_F)+\frac{m_F}{3m^2_\phi} (|A_S|^2-|A_P|^2) F^C_2(x_F)\bigg), \label{a1}
\eea
with $x_F=m^2_F/m^2_\phi$ and
\bea
F^C_1(x)&=& \frac{2}{(1-x)^4} \Big(2+3x-6x^2+x^3+6x\ln x\Big),  \\
F^C_2(x)&=&-\frac{3}{2(1-x)^3} \Big(3-4x+x^2+2\ln x\Big).
\eea
Another Yukawa type interaction of the muon is to a charged scalar $\chi$ with charge $-1$ and mass $m_\chi$ and a neutral fermion $\lambda$ with mass $m_\lambda$, given \cite{Kim:2024bub} by
\bea 
{\cal L}_2 = -{\bar \mu} \Big(C_S + C_P \gamma^5 \Big)\lambda\chi +{\rm h.c.}. \label{gen2}
\eea
In this case, we obtain the one-loop contribution to the muon $g-2$ as
\bea
a^{(2)}_\mu &=& \frac{m^2_\mu}{8\pi^2}\int^1_0 dx\, \frac{|C_S|^2 \big(x^3-x^2+\frac{m_\lambda}{m_\mu}(x^2-x)\big)+|C_P|^2 (m_\lambda\rightarrow -m_\lambda)}{(1-x)m^2_\lambda+(m^2_\chi-m^2_\mu)x+x^2 m^2_\mu} \nonumber \\
&=& \frac{m_\mu}{8\pi^2}\bigg(-\frac{m_\mu}{12m^2_\chi} (|C_S|^2+|C_P|^2 )F^N_1(x_\lambda)-\frac{m_\lambda}{6m^2_\chi} (|C_S|^2-|C_P|^2)F^N_2(x_\lambda)  \bigg), \label{a2}
\eea
with $x_\lambda=m^2_\lambda/m^2_\chi$ and
\bea
 F^N_1(x) &=& \frac{2}{(1-x)^4} \Big(1-6x+3x^2+2x^3-6x^2\ln x\Big), \\
 F^N_2(x) &=&\frac{3}{(1-x)^3} \Big( 1-x^2+2x\ln x\Big).
\eea
Using eqs.~(\ref{a1}) and (\ref{a2}), we obtain the full one-loop corrections to the muon $g-2$ due to the slepton loops \cite{Moroi:1995yh,Kim:2024bub} as
\bea
\Delta a_\mu =a^{(1)}_\mu +a^{(2)}_\mu
\eea
with
\bea
a^{(1)}_\mu &=&\sum_{j=1}^2\bigg[\frac{m^2_\mu}{96\pi^2 m^2_{\tilde\nu}} F^C_1(m^2_{{\tilde\chi}^-_j}/m^2_{\tilde\nu})\Big(g^2 |U^*_{R,j1}|^2+f^2_\mu |U^*_{L,j2}|^2 \Big) \nonumber \\
&&-\frac{g f_\mu m_\mu m_{{\tilde\chi}^-_j}}{48\pi^2m^2_{\tilde\nu}} F^C_2(m^2_{{\tilde\chi}^-_j}/m^2_{\tilde\nu}) \Big(U^*_{R,j1}U_{L,j2}+U_{R,j1}U^*_{L,j2}\Big)\bigg], \\
a^{(2)}_\mu &=&\sum_{a=1}^2\sum_{i=1}^4\bigg[-\frac{g^{\prime 2}m^2_\mu}{192\pi^2m^2_{{\tilde\mu}^2_a}} F^N_1(m^2_{{\tilde\chi}^0_i}/m^2_{{\tilde\mu}_a})\Big(|B^{R}_{i,a} |^2+|B^{L}_{i,a} |^2 \Big) \nonumber \\
&&-\frac{ g^{\prime 2} m_\mu m_{{\tilde\chi}^0_i}}{48\pi^2 m^2_{{\tilde\mu}^2_a}} F^N_2(m^2_{{\tilde\chi}^0_i}/m^2_{{\tilde\mu}_a}) \Big(B^{R}_{i,a}B^{L*}_{i,a}+B^{R*}_{i,a}B^{L}_{i,a}\Big)\bigg].
\eea
Here, $B^{R}_{i,a}, B^{L}_{i,a}$ are the effective in the form of ${\cal L}_{\rm sleptons} \supset-\sqrt{2} g' {\bar \mu}\Big(B^{L}_{i,a} P_L + B^{R}_{i,a} P_R\Big){\widetilde\chi}^0_i {\tilde \mu}_a+{\rm h.c.}$ \cite{Kim:2024bub}, and the muon Yukawa coupling is given by $f_\mu=\sqrt{2} m_\mu/(v\cos\beta)$, so the terms with the muon Yukawa coupling can be enhanced for a large $\tan\beta$. 

As a result, we approximate the one-loop corrections to the muon $g-2$ \cite{Kim:2024bub} as
\bea
a^{(1)}_\mu&\simeq&  \frac{g^2 \,m_\mu^2 m_{{\tilde\chi}^-_2} }{24\pi^2 (M_2^2-\mu_H^2)m_{\tilde\nu}^2} (\mu_H+M_2 \tan\beta) F^C_2(m^2_{{\tilde\chi}^-_2}/m^2_{\tilde\nu}) \nonumber \\
&&- \frac{g^2\, m_\mu^2 m_{{\tilde\chi}^-_1} }{24\pi^2(M_2^2-\mu_H^2)m_{\tilde\nu}^2} (\mu_H \tan\beta+M_2) F^C_2(m^2_{{\tilde\chi}^-_1}/m^2_{\tilde\nu}), \\
a^{(2)}_\mu&\simeq &\frac{g^{\prime 2}m_\mu m_{{\tilde\chi}^0_1} }{96\pi^2}\,\sin2\theta_{\tilde\mu} \bigg(\frac{1}{m^2_{{\tilde\mu}_2}} F^N_2(m^2_{{\tilde\chi}^0_1}/m^2_{{\tilde\mu}_2})-\frac{1}{m^2_{{\tilde\mu}_1}}F^N_2(m^2_{{\tilde\chi}^0_1}/m^2_{{\tilde\mu}_1})\bigg).
\eea
Here, we note that $xF^C_2(x)$ and $xF^N_2(x)$ are positive and monotonically increasing for $x>0$.
Thus, for a large $\tan\beta$, we find that the chargino contributions ($a^{(1)}_\mu$) for the muon $g-2$ can be positive.
On the other hand, for  $\mu_H>0$, the neutralino contributions ($a^{(2)}_\mu$) to the muon $g-2$ are positive for either $\sin2\theta_{\tilde\mu} <0$ (or $m^2_{{\tilde\mu}_2}>m^2_{{\tilde\mu}_1}$) (for which the lighter smuon is $SU(2)_L$ singlet-like) or $\sin2\theta_{\tilde\mu} >0$ (or $m^2_{{\tilde\mu}_2}<m^2_{{\tilde\mu}_1}$) (for which the lighter smuon is $SU(2)_L$ doublet-like). 

\begin{figure}[t]
	\centering
	\includegraphics[width=7cm,height=5.5cm]{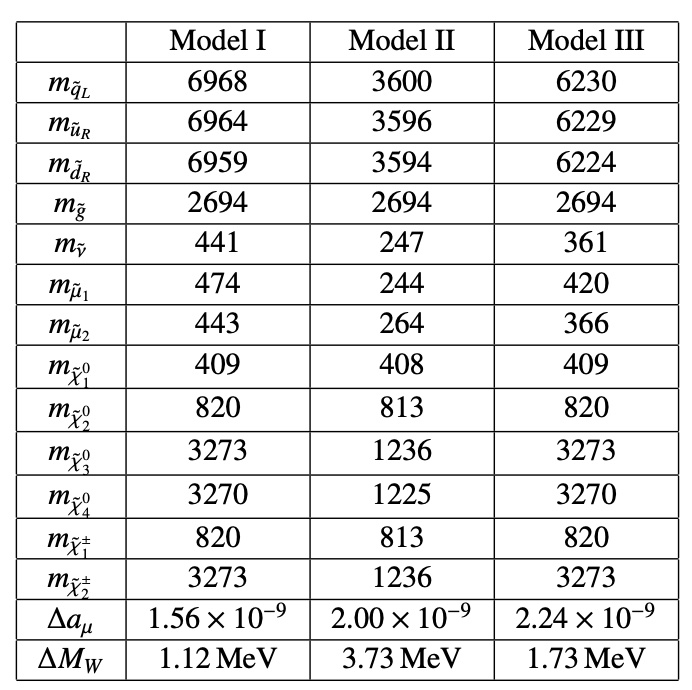} \,\,\,\,\,\,\,
	\includegraphics[width=7cm,height=5.5cm]{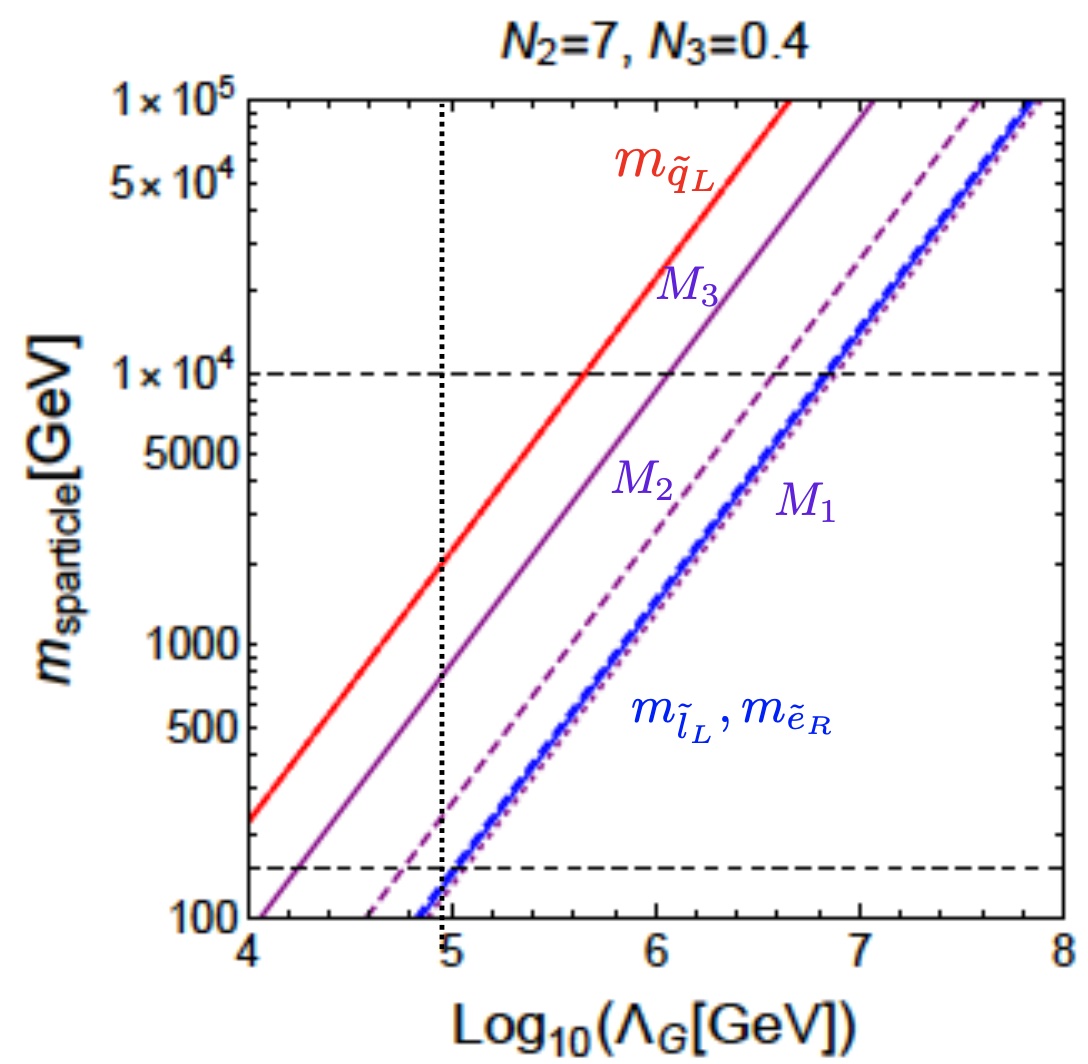}
	\caption{(Left) Superpartner masses in units of GeV and predictions for  $\Delta a_\mu$ and $\Delta M_W$  in general gauge mediation \cite{Kim:2024bub}. $\tan\beta=30$ is taken for all the models. (Right) Mass parameters for superpartners as a function of the SUSY breaking scale $\Lambda_G$ in general gauge mediation \cite{Kim:2024bub}.}
	\label{fig:masses}
\end{figure}

In the left of Fig.~\ref{fig:masses}, we present three benchmark models which are consistent with the muon $g-2$ anomaly within $1\sigma$ (Model II and III) or $2\sigma$ (Model I) in general gauge mediation \cite{Kim:2024bub}.  We listed the masses for electroweak superpartners in general gauge mediation and the predictions for $\Delta a_\mu$ and $\Delta M_W$. In particular, Model III is also consistent with the $W$ boson mass measured by CDFII within $1\sigma$. In the right plot of Fig.~\ref{fig:masses}, we also show the general-independent mass parameters for superpartners, namely, $m_{{\tilde q}_L}\simeq m_{{\tilde u}_R}\simeq m_{{\tilde d}_R}$, $M_1, M_2, M_3$, $m_{{\tilde l}_L}$ and $m_{{\tilde e}_R}$.

\begin{figure}[t]
	\centering
	\includegraphics[width=8cm,height=5cm]{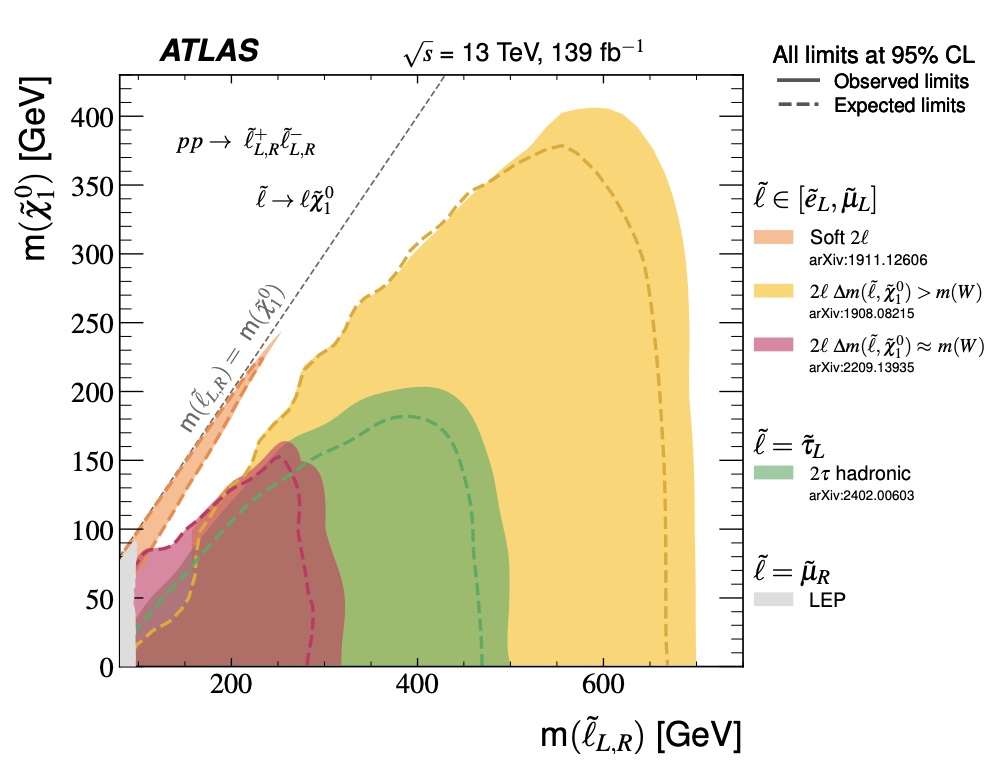} \,\,\,\,
	\includegraphics[width=7cm,height=5cm]{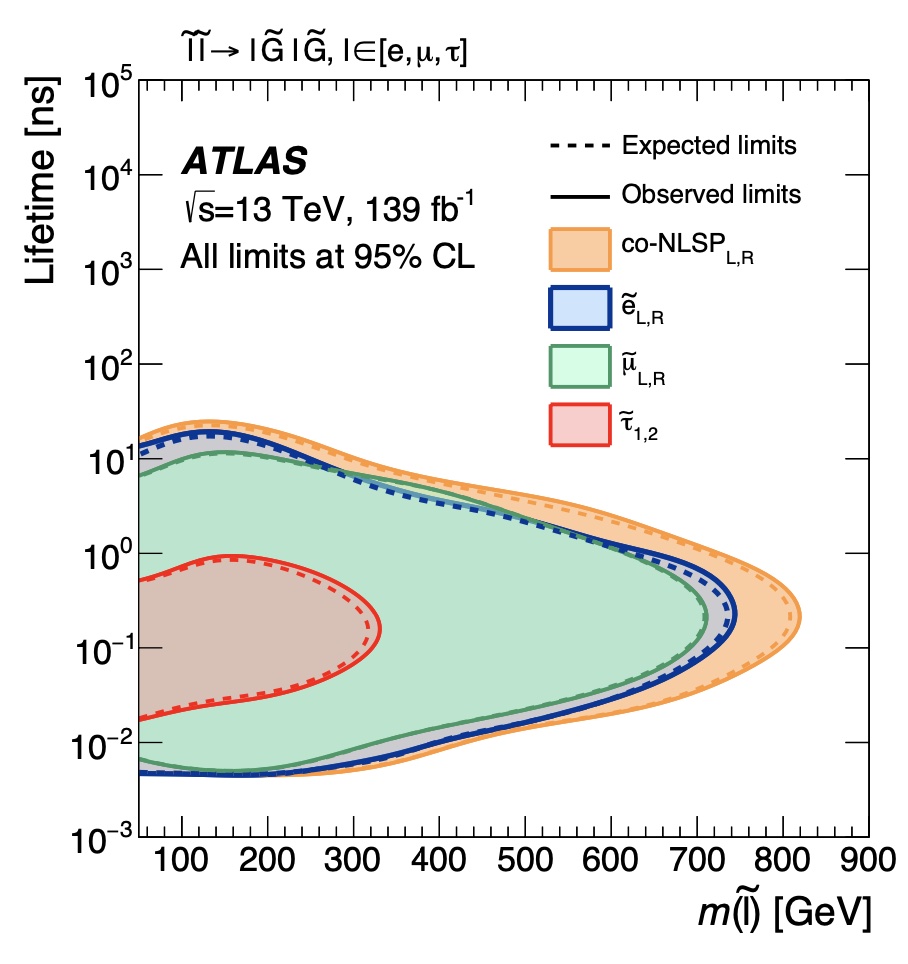}
	\caption{Searches from prompt and displaced decays of charged sleptons at LHC in the left and right plots, respectively, extracted from Refs.~\cite{ATLAS:2024lda,ATLAS:2020wjh}.}
	\label{fig:slepton}
\end{figure}

\begin{figure}[t]
	\centering
	\includegraphics[width=8cm,height=5cm]{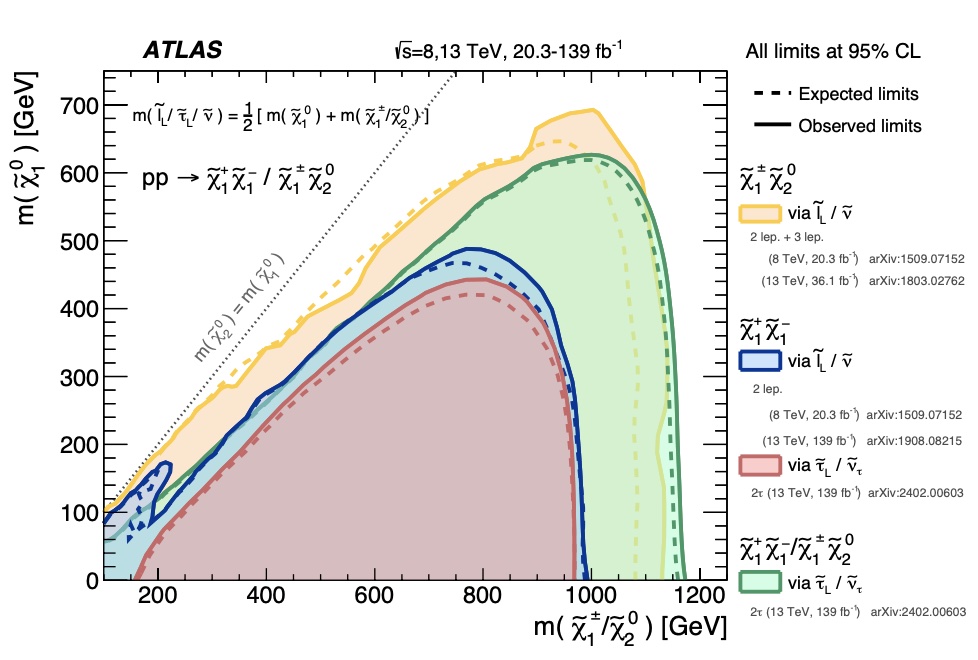} \,\,\,\,
	\includegraphics[width=7cm,height=5cm]{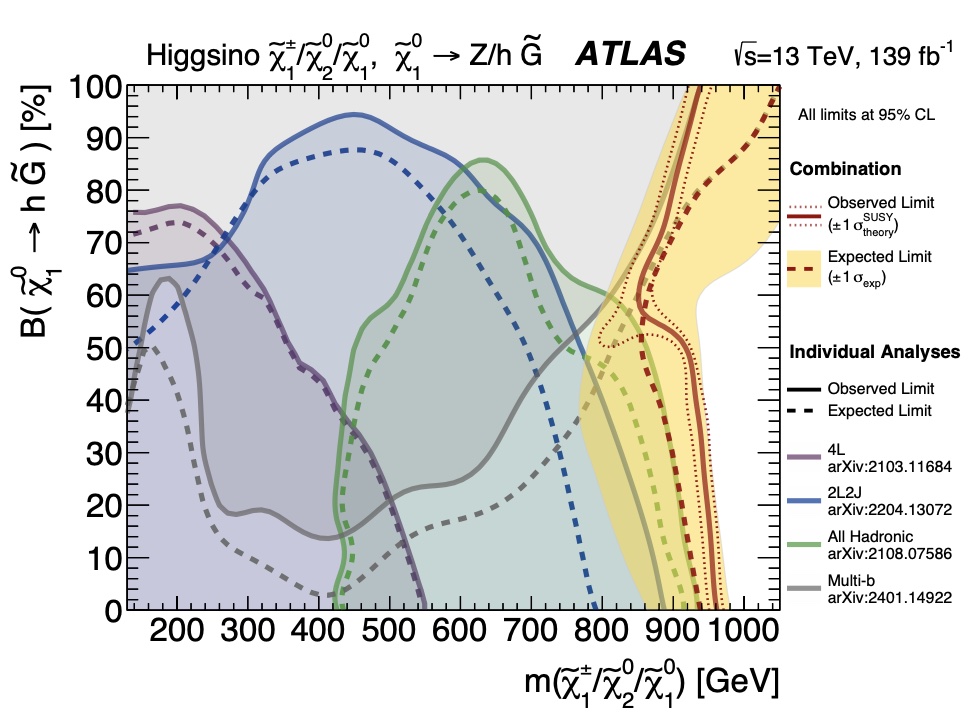}
	\caption{Searches from prompt and displaced decays of neutralinos at LHC in the left and right plots, respectively, extracted from Ref.~ \cite{ATLAS:2024lda}.}
	\label{fig:ewkino}
\end{figure}

We comment on the dark matter candidates and collider signatures with electroweak superpartners in the benchmark models.
In Model I, the neutralino is the LSP, so it can be a dark matter candidate. In this case, due to small splitting between the slepton masses and the neutralino mass, the decay products of the sleptons may be composed of soft jets or leptons \cite{ATLAS:2022hbt,ATLAS:2024lda,CMS:2024gyw} (See the left plot in Fig.~\ref{fig:slepton}). If the gravitino is lighter than the neutralino, then it can be a dark matter candidate, but displaced decays of the neutralino can lead to distinct signatures at the LHC \cite{Chakraborti:2022vds,ATLAS:2020wjh,CMS:2020bfa} (See the right plot in Fig.~\ref{fig:slepton} for the results from ATLAS).
On the other hand, in Model II and III, smuon or sneutrino is the LSP, so there is no dark matter candidate within the MSSM. 
However, if the gravitino is lighter than the sleptons, the MSSM LSP is long-lived, decaying into a pair of the gravitino and charged leptons(neutrinos) at the displaced vertex from the production point \cite{Chakraborti:2022vds,ATLAS:2024lda,CMS:2018kag,CMS:2017moi} (See the right plot in Fig.~\ref{fig:ewkino} for the results from ATLAS). 
In all the benchmark models, the heavier neutralinos or charginos have been searched for from their prompt decays, as shown in the left plot of Fig.~\ref{fig:ewkino} \cite{ATLAS:2024lda,CMS:2024gyw}. 

As can be seen in the limits for electroweak superpartners in Figs.~\ref{fig:slepton} and \ref{fig:ewkino},  the searches for prompt decays of NLSPs, which are applicable to Model I, the mass ranges up to about $700\,{\rm GeV}$ for sleptons and  about $1200,{\rm GeV}$ for chargino or heavier neutralino have been excluded. However, there is a parameter space with lower masses for electroweak superpartners still available to test by the displaced searches, which is applicable to Model II and III, so it would be interesting to keep searching for them and correlate with the indirect measurements such as muon $g-2$ and the $W$ boson mass.

\section{Supersymmetry and proton decay}

Suppose that the MSSM fields is embedded into representations in GUTs, for instance, MSSM matter fields in $\bf {\bar 5}+10$ and the MSSM Higgs fields in a pair of $5$ and ${\bar 5}$ under the SU(5) GUT.
Then, after the colored Higgsinos are integrated out, we get the effective dimension-5 operators in the superpotential, as follows,
\bea
W_5&=&\frac{\kappa}{2M_{H_C}}\, f_{u_i} f_{d_l} V^*_{kl} e^{i\varphi_i} \epsilon_{\alpha\beta\gamma}\epsilon_{rs} \epsilon_{tu} Q^{\alpha r}_i Q^{\beta s}_i Q^{\gamma t}_k L^u_l \nonumber \\
&&+ \frac{\kappa}{M_{H_C}}\, f_{u_i} e^{i\varphi_i} f_{d_l} V^*_{kl} \epsilon^{\alpha\beta\gamma} {\overline U}_{i\alpha} {\overline E}_{i} {\overline U}_{k\beta} {\overline D}_{l\gamma} \label{dim5}
\eea 
where $M_{H_C}$ is the colored Higgsino mass and the SU(5) Yukawa couplings for up-type and down-type fermions are parametrized by $y^{ij}_u=f_{u_i} e^{i\varphi_i} \delta_{ij}$ and $y^{ij}_d =V^*_{ij} f_{d_j}$ with $V_{ij}$ being the Cabibbo-Kobayashi-Maskawa(CKM) matrix. We note that $\kappa=1$ in the minimal SU(5) SU(5) GUT, but $\kappa$ can be much smaller than unity in orbifold GUT models where the tree-level Yukawa couplings for the colored Higgsinos can be forbidden at the orbifold fixed point of the extra dimension. 

The effective superpotential in eq.~(\ref{dim5}) gives rise to the effective dimension-5 interactions relevant for the proton decay between a pair of SM fermions and a pair of scalar superpartners in components,
\bea
{\cal L}_{{\rm dim-5}} &=& -\frac{\kappa}{M_{H_C}}\, f_{u_i} f_{d_l} V^*_{kl} e^{i\varphi_i} \bigg[(u_{iL} d_{iL}) ({\tilde u}_{kL} {\tilde e}_{lL})+({\tilde u}_{iL} d_{iL}) (u_{kL} {\tilde e}_{lL})- (u_{iL} d_{iL})({\tilde d}_k {\tilde \nu}_l)    \nonumber \\
&&\quad- (u_{iL} {\tilde d}_{iL})(d_k {\tilde \nu}_l) - (u_{iL} {\tilde d}_{iL})({\tilde d}_k \nu_l) - ({\tilde u}_{iL} d_{iL})({\tilde d}_k \nu_l) - ({\tilde u}_{iL} {\tilde d}_{iL})(d_k \nu_l)\bigg] \nonumber \\
&&- \frac{\kappa}{M_{H_C}}\, f_{u_i} e^{i\varphi_i} f_{d_l} V^*_{kl}  \bigg[(u^c_{iR}{\tilde e}^*_{iR} {\tilde u}^*_{kR}d^c_{lR} )+({\tilde u}^*_{iR}{\tilde e}^*_{iR} u^c_{kR}d^c_{lR} ) \bigg].
\eea
As a result,  the proton decay mode,  $p\to K^+{\bar\nu}$, appears at one-loop level with squarks and/or sleptons, so  the resultant lifetime of the proton depends on squark and slepton masses as well as chargino masses.
The effective dimension-6 operators for proton decay, $p\to K^+ {\bar \nu}_j$, with $j=2,3$, as follows \cite{Hisano:2013exa,Hisano:2022qll,Kim:2024bub},
\bea
{\cal L}_{\rm dim-6} &=& \frac{\kappa\alpha^2_2}{M_{H_C}m^2_W \sin2\beta } \bigg[\sum_{i,j=2,3} 2F(M_2, m^2_{{\tilde d}_{iL}},m^2_{{\tilde u}_{iL}}) \,{\overline m}_{u_i}{\overline m}_{d_j}  V_{u_i d} V_{u_i s} V^*_{ud_j} e^{i\varphi_i}  \nonumber \\
&&\quad\times A^{(i,j)}_R \Big( (u_L d_L) (\nu_{Lj} s_L)+  (u_L s_L) (\nu_{Lj} d_L)\Big) \nonumber \\
&&+\sum_{i,j=2,3} 2F(M_2, m^2_{{\tilde u}_{iL}},m^2_{{\tilde e}_{jL}})A^{(1,j)}_R \,{\overline m}_{u}{\overline m}_{d_j}  V_{u_i s} V^*_{u_i d_j} e^{i\varphi_i}  (u_L d_L) (\nu_{Lj} s_L) \nonumber \\
&& -\frac{{\overline m}^2_t {\overline m}_\tau V^*_{tb} e^{i\varphi_1}}{m^2_W \sin2\beta}\, F(\mu_H, m^2_{{\tilde t}_R},m^2_{{\tilde \tau}_R})\,\overline {A_R} \Big({\overline m}_d V_{ud} V_{ts} (u_R d_R)(\nu_\tau s_L)  \nonumber \\
&&\quad\quad+ {\overline m}_s V_{us} V_{td} (u_R s_R)(\nu_\tau d_L)\Big) \bigg] \label{dim6}
\eea
where $F(x,y,z)$ is the loop function, given \cite{Kim:2024bub} by
{\small
\bea
F(x,y^2,z^2) = \frac{x}{(x^2-y^2)(y^2-z^2)(z^2-x^2)}\,\bigg[ x^2 y^2 \ln\Big(\frac{x^2}{y^2}\Big)+y^2 z^2\ln\Big(\frac{y^2}{z^2}\Big)+z^2x^2\ln\Big(\frac{z^2}{x^2}\Big)\bigg], \label{protonloop}
\eea
}
$A^{(i,j)}_R, \overline {A_R} $ are the renormalization factors, and ${\overline m}_{u_i}, {\overline m}_{d_i}$ are the running quark masses defined in the $\overline{{\rm DR}}$ scheme at the scale of $\mu=2\,{\rm GeV}$ \cite{Hisano:2013exa,Hisano:2022qll}.

When we take the squark and slepton masses to be flavor-diagonal for no FCNC as in gauge mediation, we would get $m^2_{{\tilde s}_L}=m^2_{{\tilde b}_L}$ and $m^2_{{\tilde c}_L}=m^2_{{\tilde t}_L}$ between quark flavors, and $m^2_{{\tilde\tau}_R}=m^2_{{\tilde\mu}_R}$ between lepton flavors. Moreover, the electroweak symmetry relates $m^2_{{\tilde s}_L}=m^2_{{\tilde c}_L}$ and $m^2_{{\tilde b}_L}=m^2_{{\tilde t}_L}$ within the same doublets. However, we don't have to set $m^2_{{\tilde t}_R}=m^2_{{\tilde\tau}_R}$ for satisfying the bounds on FCNCs. 

The first two terms in eq.~(\ref{dim6}) are suppressed by either light quark masses or small CKM mixing angles, as compared to the third term in eq.~(\ref{dim6}).  Then, for $|\mu_H|\gtrsim M_2$ and decoupled squarks \footnote{This is in contrast to Ref.~\cite{Hisano:2022qll} where sfermion masses are taken to be universal and much heavier than TeV scale.}, namely, $m_{{\tilde u}_{iL}}, m_{{\tilde d}_{iL}},m_{{\tilde t}_{R}}\gg m_{{\tilde e}_{iL}},m_{{\tilde e}_{iR}}, M_2, |\mu_H|$,  the loop function in the first term in eq.~(\ref{dim6}) is parametrically smaller than the one in the third term in eq.~(\ref{dim6}), so we can estimate the proton lifetime from Higgsino loops \cite{Kim:2024bub} as
\bea
\tau(p\to K^+{\bar \nu}) \simeq 4\times 10^{35} \,{\rm years}\times \sin^4 2\beta \,\bigg(\frac{0.1}{\overline {A_R}}\bigg)^2 \bigg(\frac{[2F(\mu_H, m^2_{{\tilde t}_R},m^2_{{\tilde \tau}_R})]^{-1}}{10^2\,{\rm TeV}}\bigg)^2\,\bigg(\frac{M_{H_C}/\kappa}{10^{16}\,{\rm GeV}}\bigg)^2.
\eea
In comparison, we note that the current experimental limits are given by $\tau(p\to K^+{\bar \nu})>6.6\times 10^{33}\,{\rm yrs}$ \cite{Takhistov:2016eqm}.

\begin{figure}[!t]
\centering
\includegraphics[width=0.40\textwidth,clip]{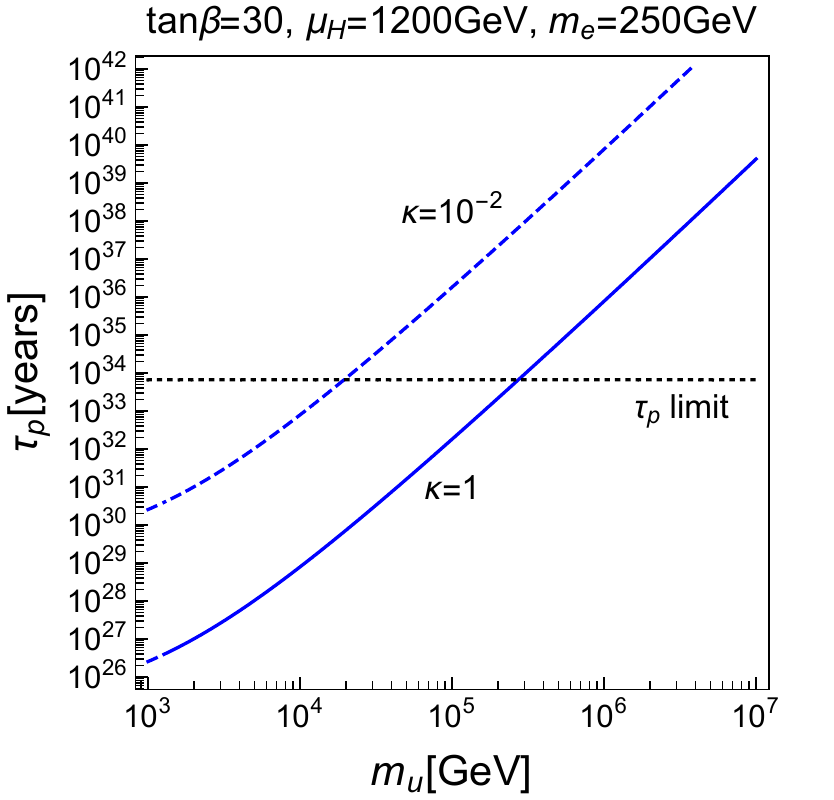}\,\,\,\,\,\,\,
\includegraphics[width=0.38\textwidth,clip]{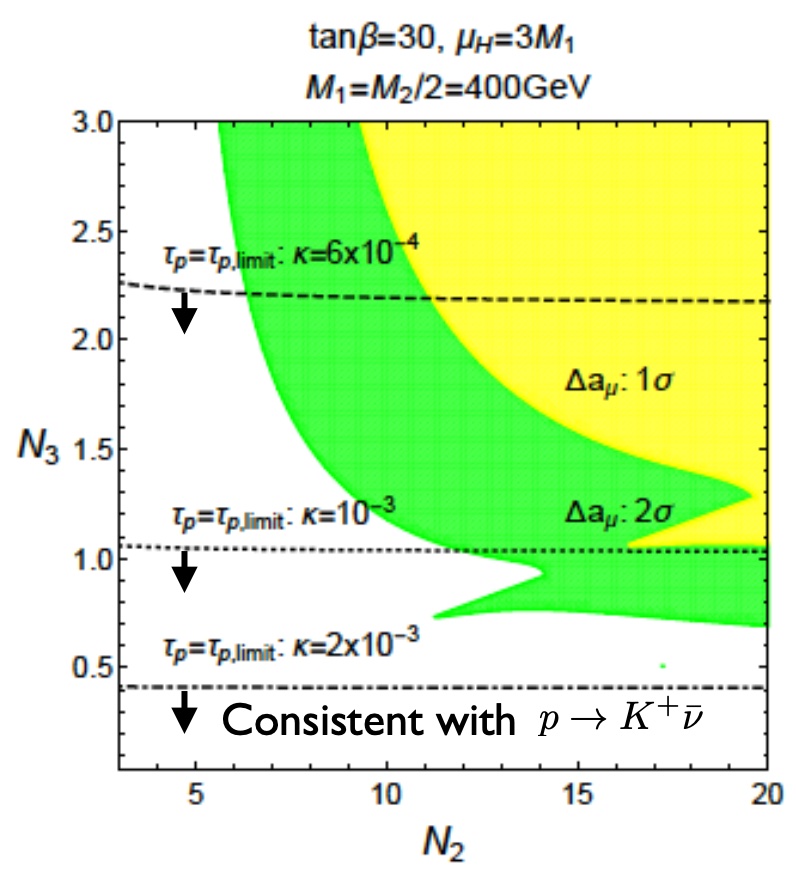} \\
\caption{(Left) Proton lifetime in years as a function of the squark mass, $m_{{\tilde u}_R}$ in units of GeV, extracted from Ref.~\cite{Kim:2024bub}. We took $\kappa=1, 10^{-2}$ in blue solid and dashed lines, respectively. We chose $\tan\beta=30$ in common, $\mu_H=1200\,{\rm GeV}$, $m_{{\tilde e}_R}=250\,{\rm GeV}$. (Right) Parameter space for the effective numbers of messenger fields in doublet and triplet, $N_2$ and $N_3$, explaining the muon $g-2$ and satisfying the bound on the proton lifetime for $p\to K^+ {\bar \nu}$, extracted from Ref.~\cite{Kim:2024bub}. The parameter space below the black lines is consistent with the bound on the proton lifetime. We chose the suppression factor, $\kappa=6\times 10^{-4}, 10^{-3}, 2\times 10^{-3}$, in dashed, dotted and dot-dashed lines, respectively. We took $\tan\beta=30$, $M_1=M_2/2=400\,{\rm GeV}$ and $\mu_H=1200\,{\rm GeV}$. 
}
\label{proton1}
\end{figure}

In general gauge mediation, we can take the squark masses to be much larger than the electroweak superpartners.
For instance, for squark masses of order $3-7\,{\rm TeV}$ and the slepton masses of order $100\,{\rm GeV}$ required for the muon $g-2$ anomaly, it is sufficient to take a suppression factor, $\kappa\sim 10^{-4}-10^{-3}$, to satisfy the bound on the proton lifetime.
From the searches for the stau sleptons at the LHC \cite{Baer:2024hgq,CMS:2022syk,ATLAS:2024fub}, we can test the flavor universality of slepton masses and get more information for the proton lifetime through $p\to K^+ {\bar \nu}$.

In the left plot of Fig.~\ref{proton1}, we obtain the proton lifetime as a function of the squark mass, $m_{{\tilde u}_R}$, in general gauge mediation. We chose the $\mu$-term and the stau mass to $\mu_H=1200\,{\rm GeV}$, $m_{{\tilde e}_R}=250\,{\rm GeV}$. We also varied the suppression factor to $\kappa=1, 10^{-2}$ in blue solid and dashed lines, respectively. The black dotted line corresponds to the current limit on the proton decay mode, $p\to K^+{\bar \nu}$. On the other hand, in the right plot of Fig.~\ref{proton1}, we overlaid the parameter space in  the effective numbers of messenger fields in doublet and triplet, $N_2$ and $N_3$, in general gauge mediation, that is favored by the muon $g-2$ at $1\sigma$ and $2\sigma$ levels, on top of the parameter space that is consistent with the proton lifetime at $95\,\%$ C.L., depending on the choice of the suppression factor.  The region below the black lines in the right plot of Fig.~\ref{proton1} is consistent with the proton lifetime. A similar discussion on the case with a large $\mu$-term and larger slepton masses for the muon $g-2$ can be referred to Ref.~\cite{Kim:2024bub}.

\section{Conclusions}
\label{sec:conclusions}

We have reviewed the framework for the minimal supersymmetric extension of the SM and the theoretical motivation for the Higgs mass, the indirect tests of superparticles from electroweak observables and the proton lifetime in the effective theory of the $SU(5)$ unification.

The current LHC bounds on colored superpartners such as gluinos and squarks are very stringent, but the SUSY explanation of the muon $g-2$ anomaly can be accommodated by relatively light electroweak superpartners such as sleptons, electroweak gauginos, Higgsinos, which can be realized in general gauge mediation.
Even if one of the sleptons is LSP in MSSM, the gravitino can be a dark matter in gauge mediation scenarios for SUSY breaking, so it would be worthwhile to keep testing this case by the displaced searches for the leptons.

In the minimal $SU(5)$ unification with TeV-scale masses for superpartners, the dimension-5 operators generated at the mass scale of the colored Higgisnos are problematic for the proton lifetime. Even in the scenarios with split masses between colored and non-colored superpartners in general gauge mediation, the problem remains because of a large decay rate of the proton coming from the stop, the light stau and the Higgsinos in loops. However, the problem of the proton lifetime can be solved when the model is embedded in the orbifold GUTs where the Higgsino wavefunctions get suppressed at the orbifold fixed point, so do the dimension-5 operators for the proton decays.

\section*{Acknowledgments}

The work is supported in part by Basic Science Research Program through the National
Research Foundation of Korea (NRF) funded by the Ministry of Education, Science and
Technology (NRF-2022R1A2C2003567).

\bibliographystyle{utphys} 

\bibliography{ref}

\providecommand{\href}[2]{#2}\begingroup\raggedright\begin{thebibliography}{10}

\bibitem{Wess:1992cp}
J.~Wess and J.~Bagger, {\em {Supersymmetry and supergravity}}.
\newblock Princeton University Press, Princeton, NJ, USA, 1992.

\bibitem{Lee:2010gv}
H.~M. Lee, S.~Raby, M.~Ratz, G.~G. Ross, R.~Schieren, K.~Schmidt-Hoberg, and
  P.~K.~S. Vaudrevange, ``{A unique $\mathbb{Z}_4^R$ symmetry for the MSSM},''
  \href{http://dx.doi.org/10.1016/j.physletb.2010.10.038}{{\em Phys. Lett. B}
  {\bfseries 694} (2011) 491--495},
  \href{http://arxiv.org/abs/1009.0905}{{\ttfamily arXiv:1009.0905 [hep-ph]}}.

\bibitem{Lee:2011dya}
H.~M. Lee, S.~Raby, M.~Ratz, G.~G. Ross, R.~Schieren, K.~Schmidt-Hoberg, and
  P.~K.~S. Vaudrevange, ``{Discrete R symmetries for the MSSM and its singlet
  extensions},'' \href{http://dx.doi.org/10.1016/j.nuclphysb.2011.04.009}{{\em
  Nucl. Phys. B} {\bfseries 850} (2011) 1--30},
  \href{http://arxiv.org/abs/1102.3595}{{\ttfamily arXiv:1102.3595 [hep-ph]}}.

\bibitem{ATLAS:2024lda}
{\bfseries ATLAS} Collaboration, G.~Aad {\em et~al.}, ``{The quest to discover
  supersymmetry at the ATLAS experiment},''
  \href{http://dx.doi.org/10.1016/j.physrep.2024.09.010}{{\em Phys. Rept.}
  {\bfseries 1116} (2025) 261--300},
  \href{http://arxiv.org/abs/2403.02455}{{\ttfamily arXiv:2403.02455
  [hep-ex]}}.

\bibitem{CMS:2019zmd}
{\bfseries CMS} Collaboration, T.~C. Collaboration {\em et~al.}, ``{Search for
  supersymmetry in proton-proton collisions at 13 TeV in final states with jets
  and missing transverse momentum},''
  \href{http://dx.doi.org/10.1007/JHEP10(2019)244}{{\em JHEP} {\bfseries 10}
  (2019) 244}, \href{http://arxiv.org/abs/1908.04722}{{\ttfamily
  arXiv:1908.04722 [hep-ex]}}.

\bibitem{CMS:2020bfa}
{\bfseries CMS} Collaboration, A.~M. Sirunyan {\em et~al.}, ``{Search for
  supersymmetry in final states with two oppositely charged same-flavor leptons
  and missing transverse momentum in proton-proton collisions at $\sqrt{s} =$
  13 TeV},'' \href{http://dx.doi.org/10.1007/JHEP04(2021)123}{{\em JHEP}
  {\bfseries 04} (2021) 123}, \href{http://arxiv.org/abs/2012.08600}{{\ttfamily
  arXiv:2012.08600 [hep-ex]}}.

\bibitem{Lee:2019zbu}
H.~M. Lee, ``{Lectures on physics beyond the Standard Model},''
  \href{http://dx.doi.org/10.1007/s40042-021-00188-x}{{\em J. Korean Phys.
  Soc.} {\bfseries 78} no.~11, (2021) 985--1017},
  \href{http://arxiv.org/abs/1907.12409}{{\ttfamily arXiv:1907.12409
  [hep-ph]}}.

\bibitem{Haller:2018nnx}
J.~Haller, A.~Hoecker, R.~Kogler, K.~M\"onig, T.~Peiffer, and J.~Stelzer,
  ``{Update of the global electroweak fit and constraints on two-Higgs-doublet
  models},'' \href{http://dx.doi.org/10.1140/epjc/s10052-018-6131-3}{{\em Eur.
  Phys. J. C} {\bfseries 78} no.~8, (2018) 675},
  \href{http://arxiv.org/abs/1803.01853}{{\ttfamily arXiv:1803.01853
  [hep-ph]}}.

\bibitem{ParticleDataGroup:2024cfk}
{\bfseries Particle Data Group} Collaboration, S.~Navas {\em et~al.}, ``{Review
  of particle physics},''
  \href{http://dx.doi.org/10.1103/PhysRevD.110.030001}{{\em Phys. Rev. D}
  {\bfseries 110} no.~3, (2024) 030001}.

\bibitem{CDF:2022hxs}
{\bfseries CDF} Collaboration, T.~Aaltonen {\em et~al.}, ``{High-precision
  measurement of the $W$ boson mass with the CDF II detector},''
  \href{http://dx.doi.org/10.1126/science.abk1781}{{\em Science} {\bfseries
  376} no.~6589, (2022) 170--176}.

\bibitem{ATLAS:2024erm}
{\bfseries ATLAS} Collaboration, G.~Aad {\em et~al.}, ``{Measurement of the
  W-boson mass and width with the ATLAS detector using
  proton\textendash{}proton collisions at $\sqrt{s}=7$ TeV},''
  \href{http://dx.doi.org/10.1140/epjc/s10052-024-13190-x}{{\em Eur. Phys. J.
  C} {\bfseries 84} no.~12, (2024) 1309},
  \href{http://arxiv.org/abs/2403.15085}{{\ttfamily arXiv:2403.15085
  [hep-ex]}}.

\bibitem{CMS:2024lrd}
{\bfseries CMS} Collaboration, V.~Chekhovsky {\em et~al.}, ``{High-precision
  measurement of the W boson mass with the CMS experiment at the LHC},''
  \href{http://arxiv.org/abs/2412.13872}{{\ttfamily arXiv:2412.13872
  [hep-ex]}}.

\bibitem{Lee:2012sy}
H.~M. Lee, V.~Sanz, and M.~Trott, ``{Hitting sbottom in natural SUSY},''
  \href{http://dx.doi.org/10.1007/JHEP05(2012)139}{{\em JHEP} {\bfseries 05}
  (2012) 139}, \href{http://arxiv.org/abs/1204.0802}{{\ttfamily arXiv:1204.0802
  [hep-ph]}}.

\bibitem{Heinemeyer:2004gx}
S.~Heinemeyer, W.~Hollik, and G.~Weiglein, ``{Electroweak precision observables
  in the minimal supersymmetric standard model},''
  \href{http://dx.doi.org/10.1016/j.physrep.2005.12.002}{{\em Phys. Rept.}
  {\bfseries 425} (2006) 265--368},
  \href{http://arxiv.org/abs/hep-ph/0412214}{{\ttfamily arXiv:hep-ph/0412214}}.

\bibitem{Muong-2:2006rrc}
{\bfseries Muon g-2} Collaboration, G.~W. Bennett {\em et~al.}, ``{Final Report
  of the Muon E821 Anomalous Magnetic Moment Measurement at BNL},''
  \href{http://dx.doi.org/10.1103/PhysRevD.73.072003}{{\em Phys. Rev. D}
  {\bfseries 73} (2006) 072003},
  \href{http://arxiv.org/abs/hep-ex/0602035}{{\ttfamily arXiv:hep-ex/0602035}}.

\bibitem{Muong-2:2021ojo}
{\bfseries Muon g-2} Collaboration, B.~Abi {\em et~al.}, ``{Measurement of the
  Positive Muon Anomalous Magnetic Moment to 0.46 ppm},''
  \href{http://dx.doi.org/10.1103/PhysRevLett.126.141801}{{\em Phys. Rev.
  Lett.} {\bfseries 126} no.~14, (2021) 141801},
  \href{http://arxiv.org/abs/2104.03281}{{\ttfamily arXiv:2104.03281
  [hep-ex]}}.

\bibitem{Muong-2:2023cdq}
{\bfseries Muon g-2} Collaboration, D.~P. Aguillard {\em et~al.},
  ``{Measurement of the Positive Muon Anomalous Magnetic Moment to 0.20~ppm},''
  \href{http://dx.doi.org/10.1103/PhysRevLett.131.161802}{{\em Phys. Rev.
  Lett.} {\bfseries 131} no.~16, (2023) 161802},
  \href{http://arxiv.org/abs/2308.06230}{{\ttfamily arXiv:2308.06230
  [hep-ex]}}.

\bibitem{Aoyama:2020ynm}
T.~Aoyama {\em et~al.}, ``{The anomalous magnetic moment of the muon in the
  Standard Model},''
  \href{http://dx.doi.org/10.1016/j.physrep.2020.07.006}{{\em Phys. Rept.}
  {\bfseries 887} (2020) 1--166},
  \href{http://arxiv.org/abs/2006.04822}{{\ttfamily arXiv:2006.04822
  [hep-ph]}}.

\bibitem{Borsanyi:2020mff}
S.~Borsanyi {\em et~al.}, ``{Leading hadronic contribution to the muon magnetic
  moment from lattice QCD},''
  \href{http://dx.doi.org/10.1038/s41586-021-03418-1}{{\em Nature} {\bfseries
  593} no.~7857, (2021) 51--55},
  \href{http://arxiv.org/abs/2002.12347}{{\ttfamily arXiv:2002.12347
  [hep-lat]}}.

\bibitem{Boccaletti:2024guq}
A.~Boccaletti {\em et~al.}, ``{High precision calculation of the hadronic
  vacuum polarisation contribution to the muon anomaly},''
  \href{http://arxiv.org/abs/2407.10913}{{\ttfamily arXiv:2407.10913
  [hep-lat]}}.

\bibitem{CMD-3:2023rfe}
{\bfseries CMD-3} Collaboration, F.~V. Ignatov {\em et~al.}, ``{Measurement of
  the Pion Form Factor with CMD-3 Detector and its Implication to the Hadronic
  Contribution to Muon (g-2)},''
  \href{http://dx.doi.org/10.1103/PhysRevLett.132.231903}{{\em Phys. Rev.
  Lett.} {\bfseries 132} no.~23, (2024) 231903},
  \href{http://arxiv.org/abs/2309.12910}{{\ttfamily arXiv:2309.12910
  [hep-ex]}}.

\bibitem{Endo:2021zal}
M.~Endo, K.~Hamaguchi, S.~Iwamoto, and T.~Kitahara, ``{Supersymmetric
  interpretation of the muon g \textendash{} 2 anomaly},''
  \href{http://dx.doi.org/10.1007/JHEP07(2021)075}{{\em JHEP} {\bfseries 07}
  (2021) 075}, \href{http://arxiv.org/abs/2104.03217}{{\ttfamily
  arXiv:2104.03217 [hep-ph]}}.

\bibitem{Kim:2024bub}
S.-S. Kim, H.~M. Lee, and S.-B. Sim, ``{Muon g-2 and proton lifetime in SUSY
  SU(5) GUTs with split superpartners},''
  \href{http://dx.doi.org/10.1103/PhysRevD.109.075035}{{\em Phys. Rev. D}
  {\bfseries 109} no.~7, (2024) 075035},
  \href{http://arxiv.org/abs/2402.04850}{{\ttfamily arXiv:2402.04850
  [hep-ph]}}.

\bibitem{Moroi:1995yh}
T.~Moroi, ``{The Muon anomalous magnetic dipole moment in the minimal
  supersymmetric standard model},''
  \href{http://dx.doi.org/10.1103/PhysRevD.53.6565}{{\em Phys. Rev. D}
  {\bfseries 53} (1996) 6565--6575},
  \href{http://arxiv.org/abs/hep-ph/9512396}{{\ttfamily arXiv:hep-ph/9512396}}.
  [Erratum: Phys.Rev.D 56, 4424 (1997)].

\bibitem{ATLAS:2022hbt}
{\bfseries ATLAS} Collaboration, G.~Aad {\em et~al.}, ``{Search for direct pair
  production of sleptons and charginos decaying to two leptons and neutralinos
  with mass splittings near the W-boson mass in $ \sqrt{s} $ = 13 TeV pp
  collisions with the ATLAS detector},''
  \href{http://dx.doi.org/10.1007/JHEP06(2023)031}{{\em JHEP} {\bfseries 06}
  (2023) 031}, \href{http://arxiv.org/abs/2209.13935}{{\ttfamily
  arXiv:2209.13935 [hep-ex]}}.

\bibitem{CMS:2024gyw}
{\bfseries CMS} Collaboration, A.~Hayrapetyan {\em et~al.}, ``{Combined search
  for electroweak production of winos, binos, higgsinos, and sleptons in
  proton-proton collisions at s=13\,\,TeV},''
  \href{http://dx.doi.org/10.1103/PhysRevD.109.112001}{{\em Phys. Rev. D}
  {\bfseries 109} no.~11, (2024) 112001},
  \href{http://arxiv.org/abs/2402.01888}{{\ttfamily arXiv:2402.01888
  [hep-ex]}}.

\bibitem{Chakraborti:2022vds}
M.~Chakraborti, S.~Iwamoto, J.~S. Kim, R.~Mase\l{}ek, and K.~Sakurai,
  ``{Supersymmetric explanation of the muon g \textendash{} 2 anomaly with and
  without stable neutralino},''
  \href{http://dx.doi.org/10.1007/JHEP08(2022)124}{{\em JHEP} {\bfseries 08}
  (2022) 124}, \href{http://arxiv.org/abs/2202.12928}{{\ttfamily
  arXiv:2202.12928 [hep-ph]}}.

\bibitem{ATLAS:2020wjh}
{\bfseries ATLAS} Collaboration, G.~Aad {\em et~al.}, ``{Search for Displaced
  Leptons in $\sqrt{s} = 13$ TeV $pp$ Collisions with the ATLAS Detector},''
  \href{http://dx.doi.org/10.1103/PhysRevLett.127.051802}{{\em Phys. Rev.
  Lett.} {\bfseries 127} no.~5, (2021) 051802},
  \href{http://arxiv.org/abs/2011.07812}{{\ttfamily arXiv:2011.07812
  [hep-ex]}}.

\bibitem{CMS:2018kag}
{\bfseries CMS} Collaboration, A.~M. Sirunyan {\em et~al.}, ``{Search for new
  physics in events with two soft oppositely charged leptons and missing
  transverse momentum in proton-proton collisions at $\sqrt{s}=$ 13 TeV},''
  \href{http://dx.doi.org/10.1016/j.physletb.2018.05.062}{{\em Phys. Lett. B}
  {\bfseries 782} (2018) 440--467},
  \href{http://arxiv.org/abs/1801.01846}{{\ttfamily arXiv:1801.01846
  [hep-ex]}}.

\bibitem{CMS:2017moi}
{\bfseries CMS} Collaboration, A.~M. Sirunyan {\em et~al.}, ``{Search for
  electroweak production of charginos and neutralinos in multilepton final
  states in proton-proton collisions at $\sqrt{s}=$ 13 TeV},''
  \href{http://dx.doi.org/10.1007/JHEP03(2018)166}{{\em JHEP} {\bfseries 03}
  (2018) 166}, \href{http://arxiv.org/abs/1709.05406}{{\ttfamily
  arXiv:1709.05406 [hep-ex]}}.

\bibitem{Hisano:2013exa}
J.~Hisano, D.~Kobayashi, T.~Kuwahara, and N.~Nagata, ``{Decoupling Can Revive
  Minimal Supersymmetric SU(5)},''
  \href{http://dx.doi.org/10.1007/JHEP07(2013)038}{{\em JHEP} {\bfseries 07}
  (2013) 038}, \href{http://arxiv.org/abs/1304.3651}{{\ttfamily arXiv:1304.3651
  [hep-ph]}}.

\bibitem{Hisano:2022qll}
J.~Hisano, ``{Proton decay in SUSY GUTs},''
  \href{http://dx.doi.org/10.1093/ptep/ptac017}{{\em PTEP} {\bfseries 2022}
  no.~12, (2022) 12B104}, \href{http://arxiv.org/abs/2202.01404}{{\ttfamily
  arXiv:2202.01404 [hep-ph]}}.

\bibitem{Takhistov:2016eqm}
{\bfseries Super-Kamiokande} Collaboration, V.~Takhistov, ``{Review of Nucleon
  Decay Searches at Super-Kamiokande},'' in {\em {51st Rencontres de Moriond on
  EW Interactions and Unified Theories}}, pp.~437--444.
\newblock 2016.
\newblock \href{http://arxiv.org/abs/1605.03235}{{\ttfamily arXiv:1605.03235
  [hep-ex]}}.

\bibitem{Baer:2024hgq}
H.~Baer, V.~Barger, and K.~Zhang, ``{Stau pairs from natural SUSY at high
  luminosity LHC},'' \href{http://dx.doi.org/10.1103/PhysRevD.110.015017}{{\em
  Phys. Rev. D} {\bfseries 110} no.~1, (2024) 015017},
  \href{http://arxiv.org/abs/2403.18991}{{\ttfamily arXiv:2403.18991
  [hep-ph]}}.

\bibitem{CMS:2022syk}
{\bfseries CMS} Collaboration, A.~Tumasyan {\em et~al.}, ``{Search for direct
  pair production of supersymmetric partners of $\tau$ leptons in the final
  state with two hadronically decaying $\tau$ leptons and missing transverse
  momentum in proton-proton collisions at $\sqrt{s}$ = 13 TeV},''
  \href{http://dx.doi.org/10.1103/PhysRevD.108.012011}{{\em Phys. Rev. D}
  {\bfseries 108} no.~1, (2023) 012011},
  \href{http://arxiv.org/abs/2207.02254}{{\ttfamily arXiv:2207.02254
  [hep-ex]}}.

\bibitem{ATLAS:2024fub}
{\bfseries ATLAS} Collaboration, G.~Aad {\em et~al.}, ``{Search for electroweak
  production of supersymmetric particles in final states with two
  \ensuremath{\tau}-leptons in $ \sqrt{s} $ = 13 TeV pp collisions with the
  ATLAS detector},'' \href{http://dx.doi.org/10.1007/JHEP05(2024)150}{{\em
  JHEP} {\bfseries 05} (2024) 150},
  \href{http://arxiv.org/abs/2402.00603}{{\ttfamily arXiv:2402.00603
  [hep-ex]}}.

\end{thebibliography}\endgroup

\end{document}